\documentclass[11pt]{article}
\pdfoutput=1
\usepackage[utf8]{inputenc}
\usepackage{dsfont}
\usepackage{diagbox}
\usepackage{amsfonts}
\usepackage{mathtools}
\allowdisplaybreaks[4]        
\usepackage{amssymb}
\usepackage{euscript}     
\usepackage{braket}
\usepackage{starfont}
\usepackage{color,soul}         
\usepackage{tensor}        
\usepackage{amsthm}
\usepackage{graphicx}
\usepackage{slashed}
\usepackage{leftidx}
\usepackage{subfigure}
\usepackage{bbm}
\definecolor{outerspace}{rgb}{0.25, 0.29, 0.3}
\definecolor{scarlet}{rgb}{1.0, 0.13, 0.0}
\usepackage[header,title,page,titletoc]{appendix}  
\definecolor{princetonorange}{rgb}{1.0, 0.56, 0.0}
\definecolor{WildStrawberry}{rgb}{1.0, 0.26, 0.64}
\definecolor{rossocorsa}{rgb}{0.83, 0.0, 0.0}
\definecolor{navyblue}{rgb}{0.0, 0.0, 0.5}
\usepackage[numbers,sort&compress]{natbib}  
\usepackage{float}
\usepackage[paper=letterpaper,margin=1in]{geometry}
\newcommand{\req}[1]{(\ref{#1})} 

\newcommand{\bea}{\begin{eqnarray}}

\newcommand{\eea}{\end{eqnarray}}
\newcommand{\ba}{\begin{eqnarray}}
\newcommand{\ea}{\end{eqnarray}}

\newcommand{\be}{\begin{equation}}
\newcommand{\ee}{\end{equation} }
\newcommand{\beqa}{\begin{eqnarray}}
\newcommand{\eeqa}{\end{eqnarray}}
\newcommand{\beqar}{\begin{eqnarray*}}
\newcommand{\eeqar}{\end{eqnarray*}}

\renewcommand{\req}[1]{eq.~(\ref{#1})}

\newcommand{\ssc}{\scriptscriptstyle}
\newcommand{\eg}{{\it e.g.,}\ }
\newcommand{\ie}{{\it i.e.,}\ }

\newcommand{\see}{S_{\ssc \rm EE}}

\DeclareMathOperator{\tr}{tr}

\usepackage{hyperref}
\hypersetup{
    colorlinks,
    citecolor=rossocorsa,
    filecolor=navyblue,
    linkcolor=navyblue,
    urlcolor=navyblue
}

\begin{document} 

\begin{titlepage}

\begin{center}

\phantom{ }
\vspace{3cm}

{\bf \Large{Reflected entropy, symmetries and free fermions}}
\vskip 0.5cm
Pablo Bueno${}^{\text{\Zeus}}$ and Horacio Casini${}^{\text{\Kronos}}$
\vskip 0.05in
\textit{Instituto Balseiro, Centro At\'omico Bariloche}
\vskip -.4cm
\textit{ 8400-S.C. de Bariloche, R\'io Negro, Argentina}



\begin{abstract}




Exploiting the split property of quantum field theories (QFTs), a notion of von Neumann entropy associated to pairs of spatial subregions has been recently proposed both in the holographic context --- where it has been argued to be related to the entanglement wedge cross section --- and for general QFTs. We argue that the definition of this ``reflected entropy'' can be canonically generalized in a way which is particularly suitable for orbifold theories --- those obtained by restricting the full algebra of operators to those which are neutral under a global symmetry group. This turns out to be given by the full-theory reflected entropy minus an entropy associated to the expectation value of the ``twist'' operator implementing the symmetry operation. Then we show that the reflected entropy for Gaussian fermion systems can be simply written in terms of correlation functions and we evaluate it numerically for two intervals in the case of a two-dimensional Dirac field as a function of the conformal cross-ratio. Finally, we explain how the aforementioned twist operators can be constructed and we compute the corresponding expectation value and reflected entropy numerically in the case of the $\mathbb{Z}_2$ bosonic subalgebra of the Dirac field.  

\end{abstract}
\end{center}

\small{\vspace{5cm}\noindent${}^{\text{\text{\Zeus}}}$pablo.bueno$@$cab.cnea.gov.ar\\
${}^{\text{\Kronos}}$casini@cab.cnea.gov.ar}

\end{titlepage}

\setcounter{tocdepth}{2}

{\parskip = .2\baselineskip \tableofcontents}


\section{Introduction}
\label{sec:Introduction}
In the context of quantum field theory (QFT), the entanglement entropy (EE) of spatial subregions is not a well-defined quantity. This is because as the cutoff is removed, more and more entanglement in ultraviolet modes across the surface is added up, leading to divergences. For the continuum model itself, the necessity of these divergences can be understood from a different perspective.  Operator algebras attached to regions are type-III von Neumann algebras. These are mathematical objects which (intrinsically) do not admit a well defined entropy --- see \eg \cite{haag,Witten:2018lha}. By the same reason, without a cutoff, a region and its complement cannot be associated with a tensor product decomposition of the Hilbert space. This tensor produt would give place to type-I factors --- the algebras of operators acting on each of the Hilbert space factors in the tensor product --- instead of type-III ones. 

Alternatively to the EE, there exist other statistical quantities that can be studied and which are finite in the continuum theory. A prototypical example is the mutual information $I(A,B)$, which, as opposed to the EE, depends on two disjoint regions $A$ and $B$ instead of one. 
The distance $\epsilon$ between the boundaries of both regions may be used as a meaningful universal regulator of EE \cite{Casini:2006ws,Casini:2015woa}, but  $I(A,B)$ remains a physical measure of correlations for arbitrary regions on its own right.     

Interestingly, in the above setting of two spatially separated regions, there is in general an intermediate tensor product decomposition of the Hilbert space separating the algebras ${\cal A}_A$ and ${\cal A}_B$ attached to those regions. This is called the ``split property'' and has been shown to hold under very general conditions controlling the growth of the number of high energy degrees of freedom  \cite{Buchholz:1973bk,Buchholz:1986dy}. 

More explicitly, a tensor product decomposition of the global Hilbert space as a product of two Hilbert spaces ${\cal H}={\cal H}_{\cal N}\otimes{\cal H}_{{\cal N}'}$ gives place to the type-I factor ${\cal N}$ corresponding to the operators acting on the first Hilbert space ${\cal H}_{\cal N}$. The split property states that there exists a decomposition where ${\cal N}$ is bigger that the algebra ${\cal A}_A$ but such that it still commutes with the operators in ${\cal A}_B$, which are included in  ${\cal N}'$. We have  
\be
 {\cal A}_A\subseteq {\cal N}\subseteq ({\cal A}_B)'\,,
\ee
where ${\cal A}'$ is the algebra of operators commuting with the algebra ${\cal A}$. It is important to note that, as opposed to  ${\cal A}_A$ or ${\cal A}_B$,  ${\cal N}$ is not the algebra of a particular geometric region. Given this structure, it is then possible to define the von Neumann entropy $S({\cal N})$ to any given split for $A$ and $B$, which is the entropy of the reduced state in one of the factors of the tensor product.   
 
While there are in general infinitely many splits associated to $A$ and $B$, there exists a particular one which can be canonically associated to a given state \cite{Doplicher:1982cv,Doplicher:1984zz,Doplicher:1983if}.\footnote{It has to be cyclic and separating for the different algebras \cite{Doplicher:1984zz}.}  The canonical type-I factor is \cite{Doplicher:1984zz} 
\be
\mathcal{N}_{AB}\equiv \mathcal{A}_A \vee J_{AB}   \mathcal{A}_A J_{AB}\,,\quad \text{or} \quad
\mathcal{N}_{AB}'=\mathcal{A}_B \vee J_{AB}   \mathcal{A}_B J_{AB}\,. \label{tomo}
\ee
In this expression $J_{AB}$ is the Tomita-Takesaki conjugation corresponding to the algebra $AB$ and the state, and ${\cal A}\vee {\cal B}$ is the algebra generated by the two algebras ${\cal A}$ and ${\cal B}$.
This  therefore defines a canonical von Neumann entropy  \cite{Longo:2019pjj}, 
\be
R(A,B)\equiv S({\cal N}_{AB})\,.
\ee
In \cite{Longo:2019pjj} this was proven to be finite for free fermions in $d=2$, and this is expected to be the case for most QFT models --- see also \cite{Narnhofer:2002ic,Otani:2017pmn,Hollands:2017dov}.

The same notion had been previously considered in \cite{Dutta:2019gen}, where it was called ``reflected entropy'' --- we shall adopt this nomenclature henceforth.\footnote{On the other hand, we use the notation ``$R(A,B)$'' to denote the reflected entropy, which differs from previous papers.} 
 This can be expressed in more simple terms for finite systems, bearing in mind that is the case of a regularized QFT.  
A state $\rho_{AB}$ defined in the Hilbert space ${\cal H}_A\otimes {\cal H}_B$
 can be purified in a canonical way as the pure state $\ket{\sqrt{\rho_{AB}}} \in (\mathcal{H}_{A} \otimes\mathcal{H}^*_{A})\otimes (\mathcal{H}_B \otimes \mathcal{H}^*_B)$.
  The reflected entropy is then defined as the von Neumann entropy associated to $\rho_{AA^*}$, which is the density matrix resulting from tracing out over $\mathcal{H}_{B}\otimes \mathcal{H}^*_B$ in the purified state. If $\rho_{AB}$ does not have zero eigenvalues, the modular conjugation operator $J_{AB}$ induced by the global pure state and the algebra ${\cal A}_{AB}$ maps precisely ${\cal A}_A$ into ${\cal A}_{A^*}$.\footnote{See for example \cite{Casini:2010nn}. For finite systems, the case of $\rho_{AB}$ with some zero eigenvalues can be dealt with by taking limits.} Then the reflected entropy coincides with the entropy of the type-I factor defined above. In particular, one has ${\cal N}_{AB}={\cal A}_{AA^*}$. By construction, the reflected entropy is a quantity depending only on ${\cal A}_{AB}$ and the state $\rho_{AB}$ in this algebra, and not on the basis chosen for the purification of this state.

 Interestingly, in \cite{Dutta:2019gen} it was shown that the reflected entropy has an expression in terms of replica manifold partition functions in QFT, giving an important practical handle for computations.  R\'enyi entropies associated to $\ket{\sqrt{\rho_{AB}}}$ can be obtained using the same expression for the R\'enyi entropy in terms of correlators involving the original fields acting on $A\cup B$ as well as those acting on $(A\cup B)^*$. 
In the same paper, the authors argued  that holographic reflected entropy can be computed from the minimal entanglement wedge cross section $E_W(A,B)$ as
\begin{equation}
R_{\rm holo.}(A,B)=2E_W(A,B)+\mathcal{O}(G_{\rm N}^0)\, ,
\end{equation} 
where $G_{\rm N}$ is Newton's constant --- see \cite{Jeong:2019xdr,Kusuki:2019rbk} for further developments. 
This updates a previous conjecture proposing that such a quantity actually equals the so-called ``entanglement of purification'' \cite{Takayanagi:2017knl,Nguyen:2017yqw}.

Some additional consequences of the  $R_{\rm holo.}(A,B)=2E_W(A,B)$ proposal  were studied in \cite{Akers:2019gcv}, where it was argued that such relation is incompatible with the previously proposed claim \cite{Cui:2018dyq} that holographic states have a mostly-bipartite entanglement structure (a similar argument in the same direction was provided assuming the entanglement of purification proposal instead). The time dependence of $R(A,B)$ on various holographic setups was studied in \cite{Kusuki:2019evw,Moosa:2020vcs,Kudler-Flam:2020url}.  Candidates for multipartite notions of reflected entropy have also been explored in \cite{Bao:2019zqc,Chu:2019etd,Marolf:2019zoo}.

In comparing reflected entropy and mutual information, we have the general inequality \cite{Dutta:2019gen}
\begin{equation}
I(A,B)\leq R(A,B)\,.\label{ini}
\end{equation}
Just like the latter, reflected entropy can also be used as a regulator of EE by letting $A^-$ be contained in some slightly greater region $(A^+)'$ (in this paper $X'$ denotes the causal complement of a region $X$). We can then define the regulator as \cite{Dutta:2019gen}
\begin{equation}
\see =\frac{1}{2}R(A^-,A^+)\,. 
\end{equation}
It can be reasonably expected that universal terms (terms that are not local and additive along the boundary of the region) should be the same when regulating with the mutual information or the reflected entropy. 

The standard split has another important application in theories with global symmetries. Let $G$ be a global symmetry group and $g\in G$. The split between $A$ and $B$ can be used to construct a {\sl twist} operator $\tau_g$ implementing the group operation in ${\cal A}_{A}$ and leaving invariant ${\cal A}_{B}$ \cite{Doplicher:1984zz}. For  
 Lie groups, Noether's theorem gives a way to construct such operators by exponentiating the local charges formed by smearing the charge density. In this sense the split allows for a version of the Noether theorem which is more general\footnote{Another advantage of the twists is that they form a representation of the group while this is not the case of the exponentials of the local smeared Noether charge.} and applies to any symmetry group \cite{Buchholz:1985ii}. 
On a different note, given a QFT and a symmetry group, we can form a new theory by considering only the operators that are invariant under such symmetry. This net of neutral operator algebras  is sometimes called the ``orbifold theory'' \cite{Dijkgraaf:1989hb}.

The first goal of the present paper is to study the reflected entropy of neutral subalgebras. We do so in Section \ref{suba}, where we point out that there exist two
 alternative definitions which extend the notion of reflected entropy to this case (reducing to it for theories without superselection sectors). One of these definitions is singled out by the simplicity of the answer: the modified reflected entropy for the subalgebra, which we call ``type-I entropy''  turns out to be the one for the original theory corrected by an explicit expression depending on the expectation value of the corresponding twist operators. 
 
Then, in Section \ref{ff}, we study Gaussian fermion systems. The standard split also gives place to a Gaussian state in this case. The reflected entropy then has a compact expression in terms of correlation functions (see also \cite{Longo:2019pjj}). This makes it amenable to numerical analysis in concrete models. We study it in detail in the case of a free massless chiral field in $d=2$ and compute the reflected entropy numerically taking the continuum limit. We also analyze the behavior of the eigenvalues of the correlator matrix for the type-I factor as the cutoff is removed and compare it with the case of the algebra of a single interval (corresponding to a type-III factor with divergent entropy in the continuum). We study how the
 standard type-I factor is distributed in the line by computing a quantity with the interpretation of a density of the algebra in terms of the fermion field operator. 

Finally, in Section \ref{z2u1} we show how twist operators for the $\mathbb{Z}_2$ fermionic and $U(1)$ symmetries of the Dirac field can be constructed, and  explicitly compute the corresponding expectation values.  From this, we compute the type-I entropy for the bosonic subalgebra using the results in Section  \ref{suba}.      
 

\section{Symmetries, twist operators, and type-I entropy}
\label{suba}
In this Section we first recall how standard splits can be used to define twist operators in theories with symmetries. Then we study possible extensions of the idea of reflected entropy for the subalgebras of operators invariant under the symmetries. This will be connected with the expectation values of twist operators.    

Let ${\cal F}$ a QFT with global internal symmetry group $G$. If we take a region $A$, the group transforms ${\cal F}_A$ into itself. But these automorphisms of ${\cal F}_A$ are outer-automorphisms, that is, they cannot be implemented by unitaries in ${\cal F}_A$. Such hypothetical unitaries would  transform  ${\cal F}_A$ while leaving the complementary algebra ${\cal F}_{A'}$ invariant. However, those transformations would be too sharply divided at the boundary of $A$ to be produced by an operator. Notwithstanding, given two spatially separated regions, $A$, $B$, there exist twist operators $\tau_g$, $g\in G$, which implement the group operation in ${\cal F}_A$ and act trivially on ${\cal F}_B$. 

Given two regions $A$, $B$, there are infinitely many possible twist operators. We consider single component disjoint regions $A$, $B$, for simplicity. As shown in \cite{Doplicher:1984zz}, an explicit standard construction 
follows  using a vector state $|\Omega\rangle$  invariant under group transformations (such as the vacuum) to produce the standard split for $A$ and $B$ explained above. The global group transformations leave the type-I factors ${\cal N}_{AB}$ and ${\cal N}_{AB}'$ in themselves. Equivalently, they act on each Hilbert space factor in the decomposition ${\cal H}_{\cal N}\otimes {\cal H}_{{\cal N}'}$ independently. This follows from (\ref{tomo}) and the fact that both ${\cal A}_{A}$ and $J_{AB}$ are invariant under the group. The latter is a consequence of ${\cal A}_{AB}$ and $|\Omega\rangle$ being invariant.   The group transformation is then implementable by a unitary $\tau_g\otimes \tau_g'$ where $\tau_g\in {\cal N}_{AB}$, $\tau_g'\in {\cal N}_{AB}'$. From this it follows that the twist operators $\tau_g$ form a representation of $G$, and they transform covariantly under the full symmetry group, 
\begin{eqnarray}
\tau_g \tau_h = \tau_{g h}\,,\quad      
g \tau_h g^{-1} = \tau_{g h g^{-1}}\,.
\end{eqnarray}

Now let us consider the orbifold theory ${\cal O}$ containing only ``neutral algebras'', \ie the operators of ${\cal F}$ invariant under $G$. We can formalize this relation with a projection $E$ of the full Hilbert space ${\cal H}_{\cal F}$ of the vacuum representation of the theory ${\cal F}$ to the one ${\cal H}_{\cal O}$ of the vacuum representation of the theory ${\cal O}$, and call with the same name the mapping of algebras $E:{\cal F}\rightarrow {\cal O}$, $E(f)=E f E$.

We would like to  obtain simple relations for the entropy in these two theories which are thus simply related to each other. These relations will be connected with the twist operators.  Any group of twists for $A$, $B$ defines a group algebra given by the linear combinations $\sum_g a_g \, \tau_g$. This algebra is isomorphic to a direct sum of full matrix algebras $\bigoplus_r M_{d_r\times d_r}$, where $d_r$ are the dimensions of the irreducible representations of $G$. This algebra has a center spanned by the projectors on each irreducible representation $r$ of $G$, corresponding to the projectors on each block in the above direct sum decomposition. These projectors can be computed from the twists as   
\be  
P_r \equiv \frac{d_r}{|G|} \sum_g \chi_r^*(g) \tau_g\,, \hspace{1cm} P_r \,P_{r'}=\delta_{r r'} P_r\,,\hspace{1cm}\sum_r P_r=1\,,\label{pro}
\ee
where  $\chi_r(g)$ is  the character of the representation $r$, and $|G|$ the order of the group. As shown in \cite{Casini:2019kex}, for the
difference of mutual informations between the two models one finds 
\begin{equation}\label{ifo}
I_{\mathcal{F}}(A,B)-I_{\mathcal{O}}(A,B) \leq - \sum_r q_r \log q_r +\sum_r q_r \log d_r^2\equiv S_\tau \,, 
\end{equation}
 where
\be
q_r\equiv \langle P_r\rangle \,,\hspace{1cm} \sum_r q_r=1\,,\label{qrr}
\ee
 are the probabilities of the different sectors of the twist group algebra, which can be computed from (\ref{pro}) in terms of the expectation values of the twists. The first term in $S_{\tau}$ is a standard entropy, whereas the second is manifestly semi-positive, which implies $S_{\tau} \geq 0$.

Therefore, (\ref{ifo}) gives us some information on the difference of mutual informations depending on expectation values of operators. This upper bound can be supplemented with a lower bound depending on expectation values of intertwiners ---  pairs of charged-anticharged operators \cite{Casini:2019kex} (see also \cite{Longo:2017mbg}). In the particular limit where $A$ and $B$ get close to touching each other, the twists expectation values tend to zero, with the exception of the identity element. In that case, (\ref{qrr}) and (\ref{pro}) give  $q_r=d_r^2/ |G|$, and the right hand side of (\ref{ifo}) becomes $\log|G|$. This is in fact the universal value of the difference of the mutual informations in the short distance limit between $A$ and $B$ \cite{Casini:2019kex,Longo:2017mbg}. See  \cite{Casini:2019kex} for the case of Lie group symmetries. This topological contribution is related to an algebraic index \cite{Longo:1989tt}. 

Now, a simple observation is that the inequality (\ref{ifo}) becomes an equation if instead of computing the mutual information between $A$ and $B$ we compute it for the standard type-I factors ${\cal{N}}$ and ${\cal N}'$, and the twists are the standard ones defined by this split. This will motivate a definition of a generalization of reflected entropy that we call ``type-I entropy'', such that the difference from the full model to the orbifold is computable in terms of twists expectation values.   

To show this let us write a basis for ${\cal H}_{\cal N}$ as $|r^{i_r},l_r\rangle$, with $i_r=1,\cdots,d_r$. For each  $r$ these vectors  transform in the index $i_r$ as the corresponding irreducible representation of the group of twists. The index $l_r$ spans  the multiplicity of the representation $r$, which is generally infinite in QFT. We define analogously $|r^{i_r\,'},l_r'\rangle$ for ${\cal H}_{{\cal N}'}$. Since the global state is pure and invariant under global group transformations it has the structure
\be
|\Omega\rangle= \sum_{r,i_r,l_r,l_{\bar{r}}'} \frac{1}{\sqrt{d_r}}   |r^{i_r},l_{r}\rangle\otimes  |\bar{r}^{\bar{i_r}},l_{\bar{r}}'\rangle  \,\, \sqrt{q_r}\, \,\alpha_{l_r,l_{\bar{r}}'}\,,
\ee
where $\bar{r}$ is the complex conjugate representation to $r$. The $q_r$ are the probabilities of the different sectors as above, and we have the normalization $\sum_{l_r,l_{\bar{r}}'}|\alpha_{l_r,l_{\bar{r}}'}|^2=1$.
Therefore, the density matrix of the system ${\cal F}$ on  ${\cal{N}}$ has the structure of a sum over blocks over the different irreducible representations
\be
 \rho^{\cal F}_{\cal N}= \bigoplus_r q_r \,\, \frac{1}{d_r} \otimes \rho_r\,, \hspace{1cm} \rho^{\cal F}_{{\cal N}'}= \bigoplus_{\bar{r}} q_{\bar{r}}' \,\, \frac{1}{d_{\bar{r}}} \otimes \rho_{\bar{r}}'\,.
\ee
In this basis, the representation of the twist group is $\bigoplus_r R_r(g)\otimes 1_r$, with $R_r(g)$ the matrices of the irreducible representation $r$. We have $q_r=q_{\bar{r}}'$, $\rho_r$ and  $\rho_{\bar{r}}'$ have the same entropy, and of course $d_r=d_{\bar{r}}$. 

The reflected entropy is 
\be
 R_{\cal F}(A,B)=S(\rho^{\cal F}_{\cal N})=\frac{1}{2} \, I_{\cal F}({\cal N},{\cal N}')=-\sum_r q_r \log q_r+\sum_r q_r \, \log (d_r)+\sum_r q_r S(\rho_r)\,,\label{dio}
\ee
where in the second equality we have used the purity of the global state. 

For the orbifold we have the neutral subalgebras $E({\cal N})$ and $E({\cal N}')$. These, however, are not type-I factors, but simply type-I algebras, because they have centers given by the projectors $P_r$ and $P_r'$ respectively, which commute with all the twists, and, as they are combinations of twists, commute with all the neutral operators in ${\cal N}$ and ${\cal N}'$ respectively. In the representation of ${\cal O}$ generated by acting with operators on the vacuum the group elements are equivalent to the identity, and therefore $\tau_g\equiv (\tau_g')^*$. This gives us $E(P_r)\equiv E(P_{\bar{r}}')$.   In this vacuum representation of the neutral algebra ${\cal O}$, the state is represented by the density matrix
\be
\rho^{\cal O}_{E({\cal N})}= \bigoplus_r q_r \, \rho_r\,, \hspace{1cm} \rho^{\cal O}_{E({\cal N}')}= \bigoplus_r q_{r} \, \rho_{\bar{r}}'\,,\hspace{1cm} \rho^{\cal O}_{E({\cal N})\vee E({\cal N}')}=\bigoplus_r q_r \, \rho_{r\bar{r}}\,,
\ee
where $\rho_{r\bar{r}}$ is pure.

Generalizing the reflected entropy (\ref{dio}) we define the type-I entropy for the orbifold as 
\be\label{ro}
S^{\rm I}_{\cal O}(A,B)\equiv\frac{1}{2} \, I_{\cal O}(E({\cal N}),E({\cal N}'))=-\frac{1}{2}\sum_r q_r \log q_r+\sum_r q_r S(\rho_r)\,.
\ee
Therefore, with this definition we have
\be
R_{\cal F}(A,B)-S^{\rm I}_{\cal O}(A,B)=\frac{1}{2}\left(- \sum_r q_r \log q_r +\sum_r q_r \log d_r^2 \right)=\frac{1}{2}S_\tau
\,.\label{mitad}
\ee
The difference between these entropies is given in terms of twist expectation values. This is exactly half the upper bound on the mutual information difference (\ref{ifo}). It follows from (\ref{ro}) and monotonicity of the mutual information that $I_{\cal O}(A,B)\le 2 S^{\rm I}_{\cal O}(A,B)$, but we cannot obtain a tighter bound as the one  (\ref{ini}) from strong subadditivity as shown in \cite{Dutta:2019gen}.

Our definition of the reflected entropy for the orbifold was motivated by simplicity of the result but we may wonder in which sense this is a natural generalization of the idea of reflected entropy previously discussed, and how it can be defined intrinsically in terms of the model ${\cal O}$ without applying to the model ${\cal F}$. This example will allow us to show that the idea of reflected entropy is richer that what one may have initially expected.

We have defined the algebra ${\cal N}_{AB}$ using (\ref{tomo}), which requires the modular conjugation of the algebra corresponding $AB$. However, for ${\cal O}$ there are two natural algebras associated to $AB$ instead of one. The algebras 
${\cal O}^1_{AB}={\cal O}_A\vee {\cal O}_B$ and ${\cal O}^2_{AB}=E({\cal F}_{AB})= ({\cal O}_{(AB)'})'$ are different, and ${\cal O}^1_{AB}\subset {\cal O}^2_{AB}$.\footnote{In dimensions $d=2 $ we have $E({\cal F}_{AB})=({\cal O}_{(AB)'})'\cap {\cal F}_{AB}$ instead.} The first one corresponds to operators generated by the neutral algebras of $A$ and $B$, while the other also contains neutral operators in $AB$ which cannot be formed by products of neutral operators in $A$ and $B$, \ie formed by charged-anticharged operators in each region.    
This failure of duality
\be
{\cal O}_{AB}\subsetneq ({\cal O}_{(AB)'})'\label{twa}\, ,
\ee
 is signalizing that ${\cal O}$ has superselection sectors given by the charged sectors of the theory (for a physical account see for example \cite{Casini:2019kex}). This is not expected to occur for complete models ${\cal F}$ without superselection sectors, that is, for models where ${\cal F}_{AB}= ({\cal F}_{(AB)'})'$.  
 
For the definition of reflected entropy in this case, we have two choices for $J_{AB}$, corresponding to the two choices of algebras. The choice of the smaller algebra ${\cal O}^1_{AB}$ coincides with the canonical choice $J_{{\cal O}_A'\cap {\cal O}_B'}=J_{{\cal O}_A\vee {\cal O}_B}$ of \cite{Doplicher:1984zz,Longo:2019pjj}, that leads to a type-I factor.
 We can still call the entropy of this factor reflected entropy $R(A,B)$.  However, not much is known on the relation of this entropy to the one of the theory ${\cal F}$. 

The second choice allows us to construct  the algebra
\be
{\cal N}_{AB}^{\cal O}={\cal O}_A\vee J {\cal O}_A J\,, \hspace{1cm} J\equiv J_{E ({\cal F}_{AB})}\,. \label{fhk}
\ee
 It follows that for any subalgebra ${\cal F}_1\subseteq {\cal F}$ we have $E J_{{\cal F}_1} E=E J_{{\cal F}_1}= J_{{\cal F}_1} E= J_{E {\cal F}_1 E}$ \cite{Doplicher:1984zz}. From this we have $(E {\cal F}_1 E)'=E {\cal F}_1' E$. In particular, $J$
  is the restriction of the modular conjugation in ${\cal F}$ to the invariant subalgebra
\be
J= J_{E({\cal F}_{AB})}= E J_{{\cal F}_{AB}}E=E J_{{\cal F}_{AB}}= J_{{\cal F}_{AB}}E\,.
\ee
Since  ${\cal O}_A=E {\cal F}_A E$ we have
\bea
&& {\cal N}_{AB}^{\cal O}= E {\cal F}_A E \vee E J_{{\cal F}_{AB}} {\cal F}_A J_{{\cal F}_{AB}} E
=((E {\cal F}_A E)' \cap (E J_{{\cal F}_{AB}} {\cal F}_A J_{{\cal F}_{AB}} E)')'\nonumber \\
&& =((E {\cal F}_A' E) \cap (E (J_{{\cal F}_{AB}} {\cal F}_A J_{{\cal F}_{AB}})' E))'= ((E {\cal F}_A'E) \cap  (E ({\cal F}_A\vee J_{{\cal F}_{AB}} {\cal F}_B J_{{\cal F}_{AB}}\vee {\cal F}_B) E)'\nonumber \\
&& =(E (J_{{\cal F}_{AB}} {\cal F}_B J_{{\cal F}_{AB}}\vee {\cal F}_B) E)'
= E ( (J_{{\cal F}_{AB}} {\cal F}_B J_{{\cal F}_{AB}}\vee {\cal F}_B))'E = E ({\cal N}_{AB}^{\cal F})\,.\label{21}
\eea
This is a type-I algebra though it is not a factor, since it has a center. This center coincides with the center of the twist algebra. We have lost the type-I factor property but a type-I algebra has a well defined entropy. Our definition of the type-I entropy for orbifolds is then a generalization of the ordinary reflected entropy, and is given by half the mutual information between this subalgebra and the one corresponding to $B$
\be
S^{\rm I}(A,B)=\frac{1}{2}I({\cal N}^{\cal O}_{AB},{\cal N}^{\cal O\,'}_{AB}) \,,\label{poe}
\ee
with ${\cal N}^{\cal O}_{AB}$ computed with (\ref{fhk}).  This coincides with (\ref{ro}). For models without superselection sectors it coincides with the usual reflected entropy
\be
S^{\rm I}_{\cal F}(A,B)=R_{\cal F}(A,B)\,.
\ee
 For orbifold theories it has the simple relation \req{mitad} with the reflected entropy of the complete model.

It is interesting to note that the limit where $A$ and $B$ touch each other, (\ref{mitad}) gives us only half the topological correction corresponding to the mutual information, $\Delta S^{\rm I}(A,B)=S^{\rm I}_{\cal F}(A,B)-S^{\rm I}_{\cal O}(A,B)=1/2 \,\log |G|$, instead of $\Delta I(A,B)=\log |G|$. This is heuristically explained as follows. For a finite system  $A B$ gets purified with the addition of $A'B'$. In the limit when $AB$ is pure the reflected  entropy $S(AA')$ duplicates the entropy of $A$ since $AB$ and $A'B'$ are decoupled. This coincides with the mutual information $I(A,B)$, which is twice the EE of $A$. However, for the orbifold, there is only one symmetry group and one center for $AA'$ which does not get decoupled, even if the states decouple. The topological part of the entropy measures precisely the non extensivity of the algebras. Thinking in comparing different regularizations of the entropy obtained with the mutual information or the reflected entropy, this curiosity may be interpreted as that there are some universal features of the entropy (produced by superselection sectors) which is possible to unambiguously distinguish with the choice of regularization.

There are other related quantities that could be defined in the context of intermediate type-I algebras. For example, we could use $S({\cal N}^{\cal O}_{AB})$ (with ${\cal N}^{\cal O}_{AB}$ given by (\ref{21})) instead of half the mutual information in (\ref{poe}). This again will lead to the reflected entropy in the case of a complete model. We  get for the entropy difference between models in this case 
\be
S({\cal N}^{\cal F}_{AB})-S({\cal N}^{\cal O}_{AB})=\sum_r q_r \log(d_r)\,.    
\ee
Curiously, this ``non Abelian" entropy is different from zero (and positive) only for non Abelian groups, where some $d_r>1$. 
Another quantity was defined in \cite{Longo:2019pjj} and called the minimal type-I entropy, which is the minimal entropy among all intermediate type-I algebras.  

To summarize, for a general theory ${\cal A}$ we can define the reflected entropy $R(A,B)=\frac{1}{2}I({\cal N}_{AB},{\cal N}_{AB}')$, where ${\cal N}_{AB}={\cal A}_A\vee J{\cal A}_A J$, and $J$ is the modular reflection corresponding to ${\cal A}_A \vee {\cal A}_B$, and the type-I entropy, given by the same formula except that $J$ is the modular reflection for ${\cal A}_{(AB)'}'$. These two coincide for complete models without superselection sectors but are different in general.

\section{Reflected entropy for free fermions}\label{ff}
In this Section we study the reflected entropy for Gaussian fermion  systems. First, we show that the reflected entropy can be obtained --- similarly to the usual entanglement entropy --- from a matrix of two-point correlators of the fermionic fields. Then, we consider the case of a free chiral fermion in $d=2$ and numerically evaluate the reflected entropy for two intervals $A$ and $B$ as a function of the conformal cross-ratio. We compare the result with the holographic one obtained in \cite{Dutta:2019gen}. We also analyze the spectrum of eigenvalues of the correlators matrix in the case of the reflected entropy and compare it to the one corresponding to a usual type-III entanglement entropy for a single interval. As we increase the number of lattice points (taking the continuum limit), the finiteness of $R(A,B)$ follows from the fact that the eigenvalues of the correlator quickly tend to fixed values. Only few of them are responsible for most of the entropy, while most eigenvalues give exponentially suppressed contributions. This is in contradistinction to the usual $\see$ case, for which an increasing number of eigenvalues becomes relevant as the continuum limit is approached, giving rise to the usual logarithmic divergence. We also define a density of the type-I algebra in terms of the ordinary field operator in the line that gives us a picture on how the factor is distributed in the line. 

\subsection{Purification of free fermions}

In this subsection we describe the purification and reflected entropy for free fermions. A more formal description can be found in \cite{Longo:2019pjj}. 

 Let $\rho$ be an invertible density matrix in a general quantum mechanical system of Hilbert space ${\cal H}_1$.  We can write 
 \begin{equation}\rho=\sum_p \lambda_p \vert p\rangle\langle p\vert\,,
 \end{equation}
  where $\lambda_p$ is the eigenvalue of $\rho$ corresponding to the eigenvector $\vert p\rangle$.
 Let $\vert \Omega \rangle$ be a purification of $\rho$ in the space ${\cal H}_1\otimes{\cal H}_2$, where ${\cal H}_2$ is a copy of ${\cal H}_1$. That is, $\rho=\textrm{tr}_{{\cal H}_2}\vert \Omega\rangle\langle \Omega\vert$.  We write  $\vert \Omega\rangle$ as a Schmidt decomposition in ${\cal H}_1\otimes {\cal H}_2$,
\begin{equation}
\vert \Omega \rangle = \sum_p \sqrt{\lambda_p}  \vert p \,\tilde{p}\rangle\,.\label{laba}
\end{equation}
 The orthonormal base $\{\vert\tilde{p}\rangle\}$ for ${\cal H}_2$ in (\ref{laba}) is arbitrary, and different basis correspond to different purifications of $\vert \Omega \rangle$. However, all these basis are equivalent for computing the reflected entropy.  
 
The modular conjugation $J$ is given by the anti-unitary operator 
\begin{equation}
J=\sum_{p q} \vert p \,\tilde{q}\rangle \langle q\,\tilde{p}\vert \, *\,,\label{jota}
\end{equation}
where $*$ is the complex conjugation in the basis $\{\vert p \tilde{q} \rangle\}$. 
 We have $J^2=1$, $J^*=J=J^{-1}$, $J\vert \Omega\rangle=\vert \Omega\rangle$. We also have the important property that the conjugation of an operator acting on the first factor gives place to an operator acting on the second one, 
\begin{equation}
J({\cal O}\otimes 1)J=1\otimes \bar{{\cal O}}\,.
\end{equation}
Defining 
\be
\Delta=\rho\otimes \rho^{-1}\,,
\ee
we have the Tomita-Takesaki relations
\be
J\, \Delta=\Delta^{-1} \,J\,,\hspace{1cm} J \Delta^{1/2} {\cal O}_1 | \Omega \rangle={\cal O}_1^* | \Omega \rangle \,,
\ee
for ${\cal O}_1$ and operator acting on the first factor.

Let $\psi_i$, $i=1,...,N$ be a system of fermions in a Hilbert space ${\cal H}_1$  of dimension $2^N$. We can purify a state given by a density matrix $\rho$ in this space by taking a Hilbert space ${\cal H}$ of double dimension and consider extending the fermion algebra with $N$ additional fermionic operators $\psi_i$, $i=1,...,2N$, such that $\{\psi_i,\psi^{\dagger}_j\}=\delta_{ij}$, $i,j=1,\cdots,2N$. 
The fermion number operator $F$ of the full system defines
\be
\Gamma=(-1)^F\,, \hspace{1cm} \Gamma^2=1\,,\hspace{1cm} \Gamma^*=\Gamma\,, \hspace{1cm} \Gamma \psi_i\Gamma=-\psi_i \,, 
\ee
and the unitary operator \cite{Doplicher:1969tk}
\be
Z= \frac{1-i \Gamma}{1-i}\,,\hspace{1cm} Z Z^* =1\,,\hspace{1cm} Z \psi_i Z^*=- i \Gamma \psi_i\,,\hspace{1cm} Z \psi_i \psi_j Z^*= \psi_i \psi_j\,.
\ee
Note this unitary transformation leaves the bosonic part of the algebra invariant.  
Let us assume the state $\rho$ is even, that is, it gives zero expectation value for products of odd number of fermion operators. It can be purified in the full space to a vector $|\Omega\rangle$ which is also even, 
\be
\Gamma|\Omega\rangle=Z |\Omega\rangle=|\Omega \rangle\,. \label{titi}
\ee 
 Given $|\Omega\rangle$ we obtain a modular reflection $J$ corresponding to the algebra of the first $N$ fermions. We have 
 \be
 \Gamma J \Gamma =J\,,\hspace{1cm} J Z=Z^* J\,,
 \ee
  because of (\ref{titi}). The operator $J \psi_i J$  commutes with $\psi_j$, $i,j\in \{1,\cdots,N\}$, and then it is not a fermion operator in the full space. However, it follows that    
defining the antiunitary \cite{Longo:2017mbg}
\be
\tilde{J}= Z \,J\,, \\ 
\ee
and\footnote{The factor $-i$ is a convenient choice of an arbitrary phase factor in this definition.}
\be
\tilde{\psi_i}^\dagger=-i \tilde{J} \psi_i \tilde{J}^* \,,\hspace{1cm}i=1,\cdots,N\,,
\ee
it follows from the algebra that the set $\{\psi_1,\cdots,\psi_N,\tilde{\psi}_1,\cdots,\tilde{\psi}_N\}$ forms a canonical anti-commutation algebra in the full space. 

The fermion correlators depend only on the density matrix $\rho$ for the first $N$ fermions. Writing for notational convenience $\psi_i^0\equiv \psi_i$, $\psi_i^1\equiv \psi_i^*$, and analogously for $\tilde{\psi}^a_i$, $a=0,1$, we have
\bea
&& \hspace{-.9cm} \langle \Omega | \psi^{a_1}_{i_1}\cdots \psi^{a_k}_{i_k}   \tilde{\psi}^{b_1}_{j_1}\cdots \tilde{\psi}^{b_l}_{j_l}
|\Omega \rangle = (-1)^{\sum b_l}\,i^l\,\langle \Omega | \psi^{a_1}_{i_1}\cdots \psi^{a_k}_{i_k}   \tilde{J} \psi^{b_1\,*}_{j_1}\cdots \psi^{b_l\,*}_{j_l}|\Omega \rangle  \label{correla}\\
&& \hspace{-.9cm} =(-1)^{\sum b_l}\,i^l\,\langle \Omega | \psi^{a_1}_{i_1}\cdots \psi^{a_k}_{i_k}   Z\, J \psi^{b_1\,*}_{j_1}\cdots \psi^{b_l\,*}_{j_l}
|\Omega \rangle=(-1)^{\sum b_l}\,i^l\,\langle \Omega | \psi^{a_1}_{i_1}\cdots \psi^{a_k}_{i_k}   Z\, \Delta^{1/2} \psi^{b_l}_{j_l}\cdots \psi^{b_1}_{j_1}
|\Omega \rangle \nonumber \\
&& \hspace{-.9cm} =(-1)^{\sum b_l}\,i^l\,i^{F_l} \langle \Omega | \psi^{a_1}_{i_1}\cdots \psi^{a_k}_{i_k}  \, \Delta^{1/2} \psi^{b_l}_{j_l}\cdots \psi^{b_1}_{j_1}
|\Omega \rangle=(-1)^{\sum b_l}\,i^l\,i^{F_l} \tr \left( \rho^{1/2} \psi^{a_1}_{i_1}\cdots \psi^{a_k}_{i_k}  \, \rho^{1/2} \psi^{b_l}_{j_l}\cdots \psi^{b_1}_{j_1}\right)\,,\nonumber
\eea
where $F_l=\frac{1+(-1)^{l+1}}{2}$ is the fermion number of $\psi^{b_l}_{j_l}\cdots \psi^{b_1}_{j_1}$.

Let us consider a Gaussian state for the fermions $\{\psi_1,\cdots,\psi_N\}$ with density matrix
\begin{equation}
\rho=\left(\det (1+e^{-K})\right)^{-1} \, e^{-\sum_{ij}\psi_i^\dagger K_{i j} \psi_j}\,,\label{dm}
\end{equation} 
for some Hermitian matrix $K$.
The two point function then fully determines the state of the system. It is given by
\begin{equation}
D_{ij}=\textrm{tr} (\rho\, \psi_i \psi_j^\dagger)=\left(\left(1+e^{-K}\right)^{-1}\right)_{ij}\,.\label{ven}
\end{equation}  
The equation (\ref{ven}) implies that $D$ is a Hermitian positive matrix with eigenvalues in $(0,1)$.
Diagonalizing $K$ we can write the density matrix as a product of thermal density matrices for independent fermion degrees of freedom
\begin{equation}
\rho=\bigotimes_k \,(1+e^{-\epsilon_k})^{-1}\, e^{-\epsilon_k c^\dagger_k c_k}\,,\label{version}
\end{equation}
with
\begin{equation}
U K U^\dagger=\epsilon=\textrm{diag}(\epsilon_1,...,\epsilon_N)\,, \hspace{1cm} c_k=\sum_l U_{kl} \psi_l\,,\hspace{.5cm} \{c_i^\dagger,c_j\}=\delta_{ij}\,,\label{uu1}
\end{equation}
and $U$ a unitary matrix. Analogously, we can define mode operators for the $\tilde{\psi}_i$ fermions with the same formula
\be
\tilde{c}_k=\sum_l U_{kl} \tilde{\psi}_l\,.\label{uu2}
\ee

From (\ref{correla}) it follows that the purified state $|\Omega\rangle$ is also a  Gaussian state for the full system of $2N$ fermions. It can be easily checked the state defined by  (\ref{correla}) is a tensor product in $k$ of states for each pair of modes $c_k, \tilde{c}_k$, and that it is Gaussian for each $k$. Then it is Gaussian for the linear combinations defined by (\ref{uu1}) and (\ref{uu2}).

We organize the fermion operators  in a single fermion field and write $\Psi_i=\psi_i$, $i=1,...,N$ and $\Psi_{i+N}=\tilde{\psi}_i$, $i=1,...,N$.   The only non zero two point correlation function is  
\begin{equation}
C_{ij}=\langle \Omega \vert \Psi_i \Psi_j^\dagger  \vert \Omega \rangle  \hspace{.7cm} i,j=1,...,2N \,.
\end{equation}
From (\ref{correla}) we obtain a block matrix representation for $C$
\begin{equation}
C=\left(
\begin{array}{cc}
D &  \sqrt{D(1-D)}\\
 \sqrt{D(1-D)} & 1-D
\end{array}
\right)\,.\label{cij}
\end{equation}
 The correlator $C$ is a projector, $C^2=C$, $C>0$, as corresponds to a global pure state $|\Omega\rangle$. 

The analogous to a region $A$ of the original system is here a subset  $A\subseteq \{1,...,N\}$. The fermion algebra of $A$ corresponds to the algebra generated by $\{\Psi_i\}_{i\in A}$. The reflected set $\bar{A}$ is the set of indices $N+i$, where $i\in A$.  The correlator matrix in a given region $X$ of the full system is just the restriction $C_X$ of $C$ to $X$, that is, $(C_X)_{ij}=C_{i,j}$ for all $i,j\in X$.  
The entropy is a function of the correlator matrix and writes  
\begin{equation}
S(X)=-\textrm{tr}(C_X \log(C_X)+(1-C_X)\log(1-C_X))\,. \label{ghh1}
\end{equation}
The same formula (\ref{cij}) can be used directly in the continuum where the matrix $C$ is a kernel $C(x,y)$, $x,y\in X$.

\subsection{Lattice calculations}
Consider a  fermionic quadratic Hamiltonian on a lattice 
\begin{equation}\label{fermiM}
H=\sum_{i,j} \psi_i^{\dagger} M_{ij} \psi_j\, ,
\end{equation}
where the fermionic operators satisfy the usual anticommutation relations $\{\psi_i,\psi_j^{\dagger} \}=\delta_{ij}$. Let $\{d_k\}$ be the basis of operators which diagonalizes $H$, namely, 
\begin{equation}
H=\sum_{l} \lambda_l\, d_l^{\dagger} d_l\, , 
\end{equation}
where $d_l\equiv \sum_j V_{lj}\psi_j$ and $[V M V^{\dagger}]_{lm}\equiv \Delta_{lm}$ with $\Delta_{lm}=\lambda_l \delta_{lm}$. The vacuum state is the Dirac sea, characterized by the conditions
\begin{align}
d_l \ket{0}= 0 \quad \text{for}\quad \lambda_l>0\,  \quad \text{and} \quad
d^{\dagger}_l \ket{0}= 0 \quad \text{for}\quad \lambda_l<0\, ,
\end{align}
namely, both annihilation operators corresponding to positive-energy modes and creation operators corresponding to negative-energy modes annihilate the vacuum. From this, it follows that $\braket{0|d_l d_k^{\dagger}|0}=\delta_{lk}$ for $\lambda_l>0$ and zero otherwise. The correlators of the original fermionic operators can be then written as
\begin{equation}\label{Dijf}
D_{ij}\equiv \braket{0|\psi_i\psi_j^{\dagger}|0}=[V^{\dagger} \theta (\Delta ) V]_{ij}\, , 
\end{equation}
where $\theta(\Delta)$ is a diagonal matrix whose diagonal is filled with ones for $\lambda_l>0$ slots and zeros for the $\lambda_l<0$ ones.

Now, let us consider a free massless chiral fermion in $d=2$, which is a function of a single null coordinate $x$. The Hamiltonian is $-\frac{i}{2}\int dx\, (\psi^\dagger  \partial \psi- \partial \psi^\dagger \psi)$. We can write a discretized Hamiltonian in a one dimensional lattice as 
\begin{equation}
H=-\frac{i}{2} \sum_j \left[ \psi^{\dagger}_j \psi_{j+1}- \psi^{\dagger}_{j+1}\psi_j \right]\, ,
\end{equation}
which takes the form of \req{fermiM} with
\begin{equation}
M_{jl}=-\frac{i}{2} \left[\delta_{l,j+1}-\delta_{l,j-1} \right]\, .
\end{equation}
It is a straightforward exercise to obtain the eigenvalues and eigenfunctions of $M$. One finds
\begin{equation}
\sum_j M_{jl} \psi^{(\lambda)}_j= \sin(\lambda) \psi^{(\lambda)}_l\, , \quad \text{where} \quad \psi^{(\lambda)}_l=\frac{e^{il \lambda }}{\sqrt{2\pi}}\, , \quad \text{and} \quad \lambda \in [-\pi,\pi]\, ,
\end{equation}
where we normalized the eigenfunctions so that $\sum_{l} \psi^{(\lambda)}_l \psi^{(\lambda')\dagger}_l=\delta(\lambda-\lambda')$. The fact that the spectrum has two zeros in $\lambda=0,\pi$ means the continuum limit of this model will describe two long-wave excitations corresponding to a doubling of degrees of freedom.

Now we can write the spectral decomposition of $M$ as
\begin{equation}
M_{jl}=\int_{-\pi}^{\pi} d\lambda \sin(\lambda)\, \psi^{(\lambda)}_j \psi^{(\lambda)\dagger}_l\, ,
\end{equation}
from which we can read the explicit expression for the fermionic correlators in the lattice $D_{jl}$ using \req{Dijf}. One finds
\begin{equation}
D_{jl}=\int_{0}^{\pi} d\lambda  \psi^{(\lambda)}_j \psi^{(\lambda)\dagger}_l= \begin{cases} \frac{(-1)^{(j-l)}-1}{2\pi i (j-l)} \quad &j\neq l\, , \\ \frac{1}{2} \quad &j=l\, . \end{cases}
\end{equation}

\subsubsection{Reflected entropy}

From the above expression for $D_{jl}$, given two disjoint regions $A$, $B$, 
we can obtain the von Neumann entropy associated to $\rho_{AA^*}$ using \req{cij} and the general expression in \req{ghh1}
as follows. 
When computing the correlators $D_{jl}$, the indices $j,l$ take values on the sites belonging to the subsets defined by $V=A\cup B$. Explicitly, if we define the discretized intervals as $A\cup B=(a_1,a_1+1,\dots,b_1-1,b_1)\cup (a_2,a_2+1,\dots,b_2-1,b_2)$, then $j$ takes values
$
 j=a_1,a_1+1,\dots,b_1-1,b_1,a_2,a_2+1,\dots,b_2-1,b_2
$,
and the same for $l$. Given $(a_1,b_1)$ and $(a_2,b_2)$ as input, we can then evaluate the matrix of correlators $D_{jl}$, which produces the first block in \req{cij}.  The lower diagonal block is simply given by $\delta_{jl}-D_{jl}$. In order to obtain the off-diagonal blocks, we diagonalize $D_{jl}$. Given its eigenvalues, $\{ d_{m} \}$, we can build the diagonal matrix $\sqrt{d_m (1-d_m)} \delta_{mn}$, and transform it back to the original basis, which yields $[\sqrt{D(1-D)}]_{jl}$. We are then left with three $(b_1-a_1+b_2-a_2)\times(b_1-a_1+b_2-a_2) $-dimensional matrices corresponding to the two diagonal blocks and the off-diagonal one, respectively. In order to obtain the von Neumann entropy associated to $\rho_{AA^*}$, we need to obtain the three submatrices corresponding to the $A$ sites in each case. These correspond to the first $(b_1-a_1)\times(b_1-a_1)$-dimensional blocks in each case. If we denote the resulting pieces by $D|_{A}$, $(1-D)|_A$ and $\left.\sqrt{D(1-D)}\right|_A$, respectively, we can finally build the matrix of correlators $C_{AA^*}$ from which we can compute the entropy of $\rho_{AA^*}$ from \req{cij} as
\begin{equation}
C_{AA^*}=\left(
\begin{array}{cc}
 D|_A& \left.\sqrt{D(1-D)}\right|_A\\
\left.\sqrt{D(1-D)}\right|_A &  (1- D)|_A
\end{array}
\right)\, .\label{cija}
\end{equation}
Given this matrix, the last step is to obtain its eigenvalues, $\{ \nu_{m} \}$. Finally, the reflected entropy is given by
\begin{equation} \label{genSVaa}
R_{\rm ferm.}=-\sum_m \left[ \nu_{m} \log(\nu_{m})+(1_m-\nu_{m})\log(1_m-\nu_{m})\right]\, .
\end{equation}
In the following, when showing results for the chiral fermion, we take into account the fermion doubling by dividing the numerical results for the entropy by $2$. Note that results normalized by the central charge $(c+\bar{c})/2$ are equal for the chiral and Dirac fermion. 

\begin{figure}[!t]
	\centering 
	\includegraphics[scale=0.73]{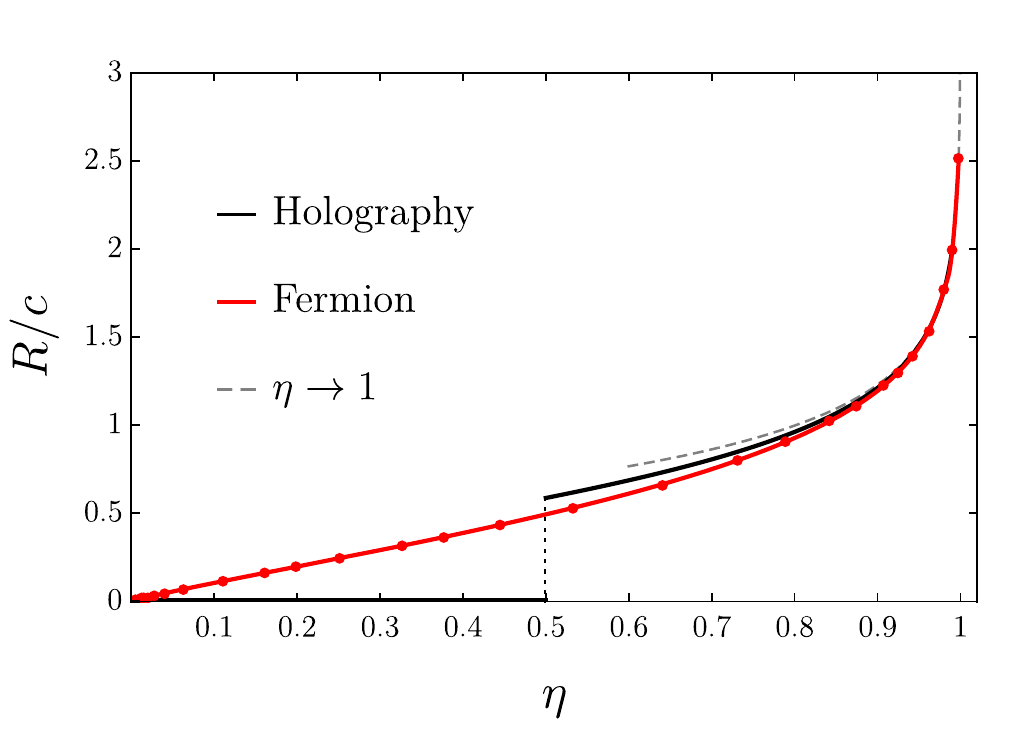}
	\caption{Reflected entropy normalized by the central charge, $R/c$, as a function of the conformal cross-ratio $\eta$ for holographic Einstein gravity (black) and a free fermion (red line and dots). The gray dashed line corresponds to the general-theory behavior for $\eta \rightarrow 1$. For $\eta=1/2$ the holographic result undergoes a phase-transition and the leading $N$ term  drops to zero for smaller values of the cross ratio. }
	\label{ratiodd}
\end{figure}

In the continuum limit this entropy should be a function $R_{\rm ferm.}(\eta)$ of the cross-ratio
\begin{equation}
\eta\equiv \frac{(b_1-a_1)(b_2-a_2)}{(a_2-a_1)(b_2-b_1)}=\frac{L_A L_B}{(d+L_A)(d+L_B)}\, ,
\end{equation}
where $L_{A,B}$ are the two interval lengths and $d$ the separating distance. 
For each value of $\eta$, obtaining the continuum-limit result for $R_{\rm ferm.}$ entails considering a sufficiently large number of points in our discretized intervals. As we increase such number with fixed $\eta$,  the results asymptotically approach certain values which correspond to the continuum ones, and we extrapolate to infinite size by a polynomial fit in the inverse size of the system.\footnote{Naturally, the number of points required to stabilize the corresponding value of  $R_{\rm ferm.}$  grows with $\eta$.}  These are the ones shown in Fig. \ref{ratiodd} 
and, as expected, they are finite for all values of $\eta$.

In Fig. \ref{ratiodd} we have also included the holographic result obtained in  \cite{Dutta:2019gen} using replica methods,
\begin{equation}
R_{\rm holo.}(\eta)= \begin{cases} 
\frac{2c}{3}\log \left[\frac{1+\sqrt{\eta}}{\sqrt{1-\eta}} \right] +\mathcal{O}(c^0)\, , \quad \text{for} \quad \eta>1/2\, , \\
\mathcal{O}(c^{0})\, , \quad   \quad \quad \quad \quad \,  \, \, \,   \, \,  \, \, \,   \, \,  \, \quad  \text{for} \quad \eta<1/2\, .
\end{cases}
\end{equation}
This in turn agrees with the $E_W$ calculations of \cite{Takayanagi:2017knl,Nguyen:2017yqw}. Normalizing by the central charge, the fermion result turns out to be remarkably close to (and always smaller than) the holographic one for all values of $\eta >1/2$. For $\eta=1/2$, the holographic result has a phase transition and the leading $c$ term drops to zero. On the other hand, the fermion one continuously goes to zero as $\eta\rightarrow 0$. 

Finally, as argued in \cite{Dutta:2019gen}, for $\eta\rightarrow 1$ the reflected entropy in a $d=2$ CFT universally behaves as
\begin{equation}
R(\eta\rightarrow 1)=-\frac{c}{3}\log (1-\eta)+\frac{c}{3}\log 4\, ,
\end{equation}
which we also included in Fig. \ref{ratiodd}. Both the holographic and fermion results approach the limiting curve from below.

For small values of $\eta$, we find that the approximation
\begin{equation}
R_{\rm ferm.}(\eta\rightarrow0)/c \sim  -0.15 \eta \log \eta +0.67 \eta+ \dots
\end{equation}
fits well the numerical data. The above expression is to be taken with a grain of salt, in the sense that including more or less numerical points in the interpolation slightly (but significantly) modifies the coefficients. However, we do seem to observe that the $\eta \log \eta $ term is required to properly account for the data. The appearance of this term is interesting when compared to the mutual information case. For that, the exact answer for a chiral fermion reads \cite{Casini:2009vk,Casini:2005rm}
\begin{equation}\label{mutuaf}
I_{\rm ferm.}(\eta)=-\frac{1}{6}\log (1-\eta)\, \quad \Longrightarrow \quad I_{\rm ferm.}(\eta\rightarrow 0)=\frac{1}{6}\sum_{j=1} \frac{\eta^j}{j}\, .
\end{equation}
Hence, in the small  $\eta$ limit, the mutual information is given by a  power law with no logarithmic corrections.

As we have mentioned above, the values shown in the figure correspond to the continuum limit. Naturally, an analogous limiting procedure in the case of the usual entanglement entropy gives rise to divergent expressions (involving the usual logarithmic term in the case of $d=2$ CFTs). If we were computing entanglement entropy, we would use the same expression as in  \req{genSVaa} where now the eigenvalues would be ones of $D|_A$. 
The reason why the same formula when applied to $D|_A$ gives rise to a divergent expression whereas it produces a finite entropy when applied to $C_{AA^*}$ may look somewhat obscure from the point of view of this lattice approach. In order to shed some light on this, we next compare the spectrum of $D|_A$ (corresponding to the usual entanglement entropy of a single interval) with the one of $C_{AA^*}$.

\subsubsection{Correlators matrix spectrum}
Both the usual entanglement entropy and the reflected entropy are von Neumann entropies. The first corresponds to a regularization of a type-III algebra associated to the corresponding entangling region $A$ (in the simplest possible a case, a single interval) whereas the second is associated to the type-I algebra canonically related to two given regions $A$ and $B$ (here, two intervals). As we saw above, this means that for Gaussian systems both quantities can be evaluated from matrices of correlators: $D|_A$ and $C_{AA^*}$ respectively, using the same formula appearing in the RHS of \req{genSVaa} --- more generally, \req{ghh1}. In the former case, the  $\nu_k$ stand for the eigenvalues of $D|_A$, and in the latter those correspond to the eigenvalues of $C_{AA^*}$. In fact, as we saw, $C_{AA^*}$ includes $D|_A$ as one of its block submatrices.

In spite of these ``similarities'', the result obtained for the reflected entropy is very different from the one corresponding to the entanglement entropy. While the former can be used as a regulator for the latter as we make both regions come close ($\eta\rightarrow 1$ above), the reflected entropy is otherwise finite for all values of the conformal cross ratio, whereas the entanglement entropy of a single interval diverges logarithmically in the continuum, $\see=\frac{c}{3} \log(L_A/\epsilon)$. This different behavior can be traced back to the properties of the respective spectra of $D|_A$ and $C_{AA^*}$. As should be clear from \req{genSVaa}, eigenvalues close to $1$ (or $0$) make little contribution to the corresponding von Neumann entropy. On the other hand, the closer to $1/2$, the greater the contribution from the corresponding eigenvalue. From this perspective, it is expectable that a finite result for the entropy should be associated to the existence of a finite number of eingenvalues significantly different from $1$, and viceversa --- \ie an infinite entropy should be related to the appearance of an increasing number of $\nu_k \not\simeq 1$ eigenvalues as we go to the continuum. 

\begin{figure}[t]
	\includegraphics[scale=0.661]{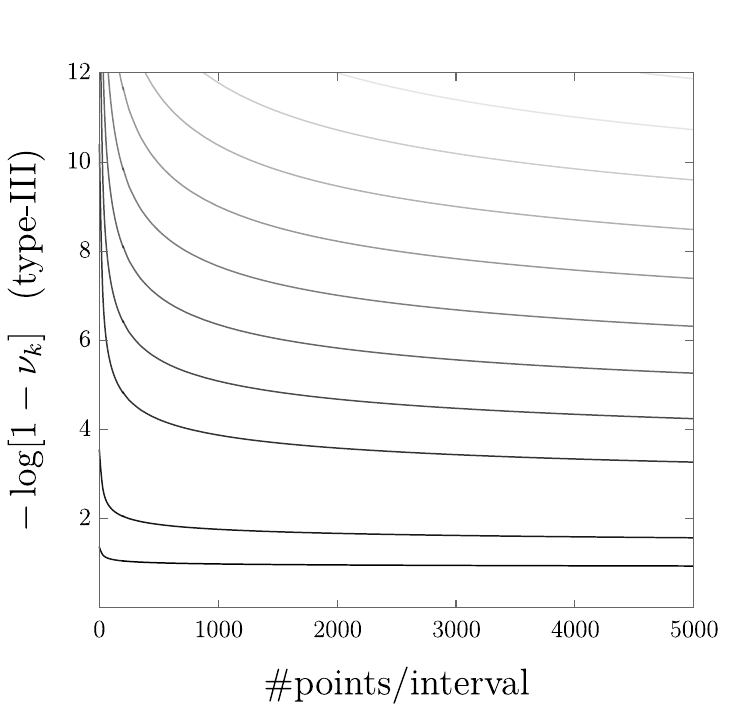}
\includegraphics[scale=0.661]{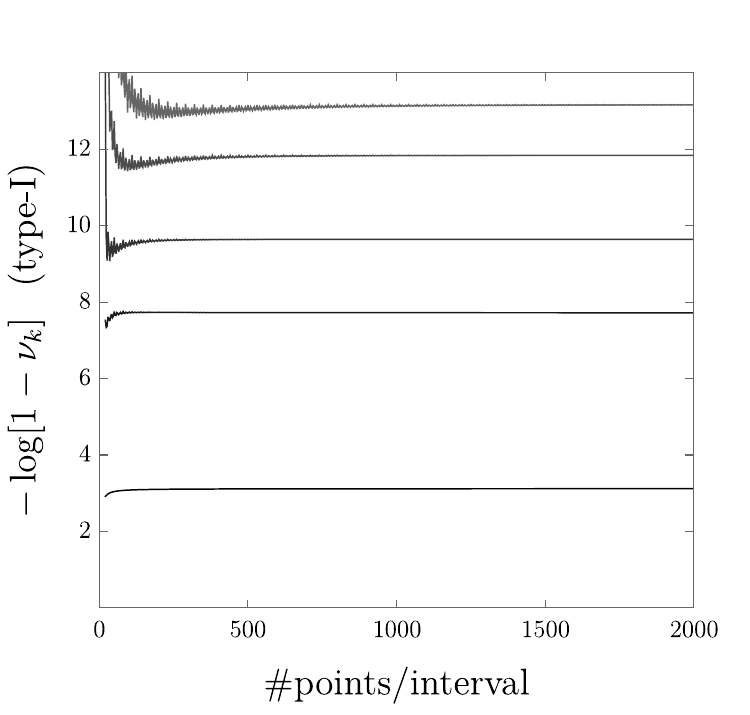}
	\caption{We plot the ``leading'' eigenvalues of $D|_A$ and $C_{AA^*}$ (as defined in the main text) corresponding, respectively to: the correlators matrix required for the evaluation of the usual type-III entanglement entropy for a single interval and the reflected entropy $R(A,B)$ for a fixed value of the cross-ratio $\eta=25/36$, for different numbers of lattice points. The plot is logarithmic to make the behavior of the different eigenvalues more visible.}
	\label{eigen}
\end{figure}

In order to analyze these features, we numerically computed the eigenvalues of $D|_A$ and 
$C_{AA^*}$ (for a fixed value of the cross-ratio, here we take $\eta=25/36$) and arranged them from closest  to farthest to $1/2$. Since the spectrum is symmetric around $1/2$ it is enough to consider the eigenvalues $1 > \nu_k\ge 1/2$. 
In each case, we refer to the ``leading'' eigenvalue as the one which is closest to $1/2$, and so on. We plot the results for the leading eigenvalues in Fig. \ref{eigen}. As we approach the continuum limit, a growing number of eigenvalues of $D|_A$ becomes relevant and separate from $1$, giving rise to the logarithmically divergent behavior.  
No such phenomenon occurs for the type-I factor, where we observe that any fixed eigenvalue quickly tends to  a constant value in the continuum limit, and only few of them are not exponentially close to $1$ as we approach that limit.
For a fixed cross ratio, a few eigenvalues are enough to account for the whole entropy in the continuum. 

The fact that $R(A,B)$ is essentially controlled by a couple of eigenvalues of $C_{AA^*}$ can be verified by defining the ``partial'' reflected entropies
\begin{equation}
 R^{(p)}_{\rm ferm.}(\eta)/c=-2\sum_{m,\,\nu_m>1/2}^p \left[ \nu_{m} \log(\nu_{m})+(1_m-\nu_{m})\log(1_m-\nu_{m})\right]\, ,
\end{equation}
where it is understood that the eigenvalues have been arranged from closest to farthest to $1/2$ and the factor $2$  comes from the fact that the eigenvalues appear mirrored with respect to $1/2$. Also, $ R^{(\infty)}_{\rm ferm.}(\eta)= R_{\rm ferm.}(\eta)$ is just the reflected entropy by definition. For instance, for $\eta=25/36\simeq0.6944$ we find
\begin{align*}
 R^{(1)}_{\rm ferm.}(25/36)/c&\simeq 0.7248\, ,\\
  R^{(2)}_{\rm ferm.}(25/36)/c&\simeq 0.7403\, ,\\
    R^{(3)}_{\rm ferm.}(25/36)/c&\simeq 0.7430\, ,\\
    R^{(4)}_{\rm ferm.}(25/36)/c&\simeq 0.7434  \, , \\
      R^{(\infty)}_{\rm ferm.}(25/36)/c&\simeq 0.7436\, .
\end{align*}
The consideration of the leading eigenvalue already provides a decent approximation to the reflected entropy. Since $R^{(p)}_{\rm ferm.}/c\le 2 p \log(2)$ however, the number of relevant eigenvalues increases logarithmically as $\eta\rightarrow 1$. 

For the case of the interval (type-III factor) the continuum limit corresponds to the correlator kernel of the fermion which has continuum spectrum covering all the interval $(0,1)$. This spectrum is given by $\nu(s)=\frac{1+\tanh(\pi s)}{2}$, in terms of a parameter $s\in (-\infty,\infty)$ having uniform density in the line \cite{Casini:2009vk}. This gives a density of eigenvalues in the variable $\nu$ given by $ds/d\nu\propto (\nu(1-\nu))^{-1}$. The integrated number of eigenvalues for $\nu>1/2$ is then proportional to $\log(\nu/(1-\nu))$, equispaced in logarithmic variable as we approach $\nu\sim 1$. This is readily seen in Fig. \ref{eigen}.

We have just analyzed how the spectra of eigenvalues of the correlators matrix differs in the case of the reflected entropy with respect to the one of a single-interval (corresponding to an usual logarithmically divergent type-III entanglement entropy). It is convenient to mention that the spectrum of reduced density matrices in the entanglement entropy context has been subject of intense study --- see \eg \cite{Chung_2000,Calabrese_2008,2013PhRvB..87w5107L,Laflorencie:2015eck,Alba:2017bgn} and references therein. In the case of free fermions (and more generally for Gaussian states), this ``entanglement spectrum'' is naturally related to the spectrum of the corresponding corelators matrix as follows.

Both the eigenvalues of the correlators matrix $\{ \nu_{k} \}$ and the density matrix $\{ \lambda^{(\rho)}_{j} \}$ can be written in terms of the ones of the Hamiltonian, $\{ \varepsilon_k\}$. Indeed, the former are related, one-to-one, to the $\{ \varepsilon_k\}$ by
\begin{equation}
\nu_{k}=\frac{1}{1+e^{-\varepsilon_k}}\, , \quad \text{and therefore:} \quad \varepsilon_k=\log \left[1/\nu_{k}-1 \right]\, .
\end{equation}
On the other hand, the eigenvalues of the density matrix are given by the set
\begin{equation}
\left\{ \lambda^{(\rho)}_{j} \right\}= \left\{ \prod_k \frac{e^{-\varepsilon_k o_k}}{1+e^{-\varepsilon_k}}\, , \, o_k\in \{0,1 \}  \right\}\, .
\end{equation}
Therefore, we can write them in terms of the $\nu_{k}$ as\footnote{For instance, for $k=1,2$, we have 
\begin{equation}
\{\lambda^{(\rho)}_{j}\} = \left\{(1-\nu_{1})(1-\nu_{2}),\nu_{2}(1-\nu_{1}),\nu_{1}(1-\nu_{2}),\nu_{2}\nu_{1} \right\}\, .
\end{equation}
Note that the number of eigenvalues of the density matrix grows exponentially with the number of lattice points.}
\begin{equation}
\left\{\lambda^{(\rho)}_{j}\right\} = \left\{ \prod_k \left[1-\nu_{k}\right]^{1-o_k} \nu_{k}^{o_k}\, , \, o_k\in \{0,1 \}  \right\}\, .
\end{equation}

The eigenvalues of the correlator $\nu_k$ are then just the probabilities in the two-dimensional density matrix of each independent fermion degree of freedom. From the point of view of the lattice calculation both the type-I and type-III factors appear in the continuum limit as  an infinite tensor product of single fermion degrees of freedom. The resulting type of von Neumann algebra depends on the state, which is necesary to define the limit of the tensor product \cite{Witten:2018lha}. This state is given by the probabilities $\nu_k$ for each mode. A sure sufficient condition of the result being a type-I algebra is that the sum of the entropies of the different modes converges. If it does not, different results may be obtained according to the behavior of the sequence of $\nu_k$ in the continuum limit. See \cite{Witten:2018lha} for examples where the limit is a type-III$_\lambda$ factor for $\lambda \in [0,1]$. The case of the algebra of the interval is known to be a type-III$_1$ factor which requires that the $\nu_k$ have at least two accumulation points in $(0,1)$. As we have seen, the fermion field correlator in the interval has indeed a continuum spectrum in $(0,1)$, and all points are accumulation points, proving that is a type-III$_1$ factor. This accumulation of eigenvalues in any point is also visible numerically from Fig. \ref{eigen}.

\subsection{Spatial density of the standard type-I factor}
As opposed to the usual type-III algebras associated to subregions, the type-I factor ${\cal N}_{AB}$ cannot be sharply associated to any region.  In order to make this heuristic observation more precise, we can define a notion of ``spatial density'' which measures how ${\cal N}_{AB}$ is distributed in the line.

Suppose first that we have two sets of fermion fields linearly related to each other 
\be
\phi(u)=\int dv \,  K(u,v)\,  \tilde{\phi}(v)\,.
\ee
Anti-commutation relations for both fields
\be 
\{\phi(u),\phi^\dagger(u')\}=\delta(u-u')\,,\hspace{1cm} \{\tilde{\phi}(v),\tilde{\phi}^\dagger(v')\}=\delta(v-v')\,,
\ee 
imply
\be
K(u,v)=\{\phi(u),\tilde{\phi}^\dagger(v)\}\,, \hspace{1cm} \int dv\,K(u,v)K(u',v)^*=\delta(u-u')\,.
\ee
Now consider a subalgebra of the fermion system generated by the fields  $\tilde{\phi}(v)$ in a subset of the line, $v\in V$. We would like to understand how this subalgebra is distributed in the line of coordinate $u$. A natural density $d(u)$ is given by 
\be
 \int_V dv\,K(u,v)K(u',v)^*\sim d_V(u)\, \delta(u-u')\,.
\ee
This tells us the proportion of the field $\phi(u)$ that can be reconstructed from the algebra in $V$. We have in particular 
\be
0\le d_V(u)\le 1\,,\hspace{.7cm} \sum_j d_{V_j}(u)=1\,, \hspace{.4cm}\textrm{for} \,\, \cup_j V_j=W\,, i\neq j\implies V_i\cap V_j=\emptyset \,,  
\ee
where we have denoted by $W$ the full domain of the variable $v$. 

As anticipated, we now apply this idea to understand how  the  type-I  factor ${\cal N}_{AB}$ is distributed along the line. One set of fields is given by 
\be 
\{\psi(y), y\in A\}\cup \{\tilde{\psi}(y)=i\tilde{J}_{AB}\psi^\dagger(y) \tilde{J}_{AB}^*, y\in A\}\,, \label{espress}
\ee
which spans ${\cal N}_{AB}$. This is completed by ${\cal N}_{AB}'$ which is generated by the same expression (\ref{espress}) but where $y\in B$. We need to determine the density of this set in terms of 
the fields $\psi(x)$, $x\in R$. Then we have to compute
\be
\int_A dy \, \{\psi(x),\psi^\dagger(y)\}\{\psi(x'),\psi^\dagger(y)\}^*+\int_A dy \, \{\psi(x),\tilde{\psi}^\dagger(y)\}\{\psi(x'),\tilde{\psi}^\dagger(y)\}^* \,,\label{dos}
\ee
and look for the $\delta (x-x')$ term. 
It is evident that $d(x)=1$ for $x\in A$, and $d(x)=0$ for $x\in B$. The density for $x\in (AB)'$ is determined by the second term in (\ref{dos}). We write this term using 
\bea
 \{\psi(x),\tilde{\psi}^\dagger(y)\} &=& \langle  \Omega|\{\psi(x),\tilde{\psi}^\dagger(y)\}|\Omega \rangle=-i \langle  \Omega|\psi(x)\tilde{J}\psi(y)|\Omega \rangle-i \langle  \Omega|\psi^\dagger(x)\tilde{J}\psi^\dagger(y)|\Omega \rangle^*\\ &=&2\,i\, \textrm{Im}\,\langle  \Omega|\psi(x)\Delta^{1/2} \psi^\dagger(y)|\Omega \rangle\,,\nonumber
\eea
where $x\in (AB)'$, $y\in A$. 

\begin{figure}[!t]
	\centering 
	\includegraphics[scale=0.75]{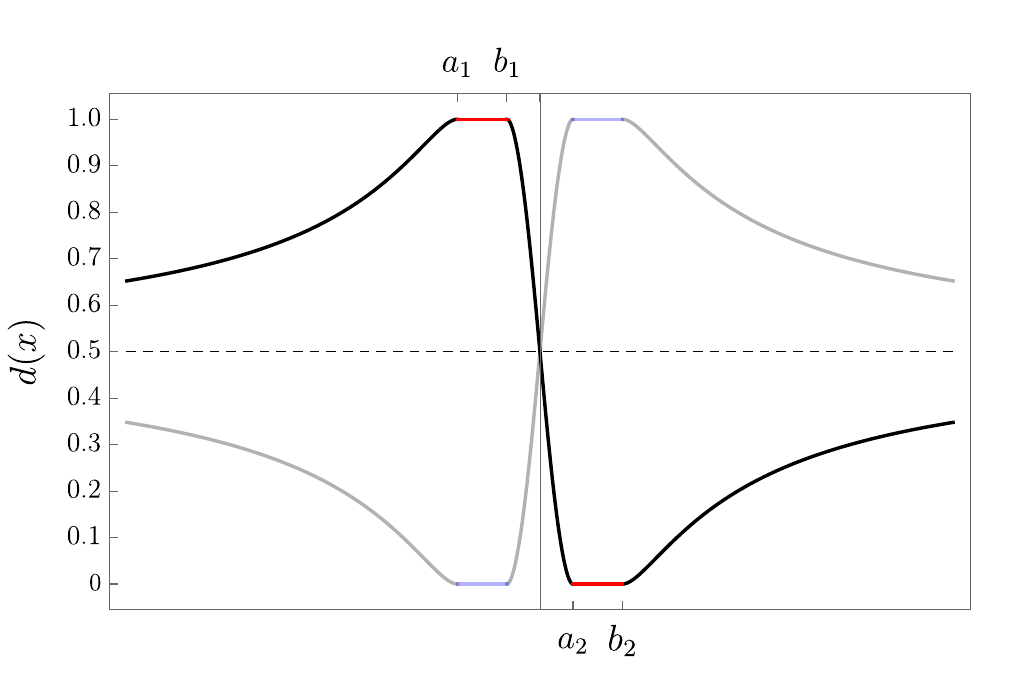}
	\caption{Spatial density $d(x)$ for the type-I factor ${\cal N}_{AB}$ (black curve and red lines). For $x \in A$, $d(x)=1$, whereas for $x\in B$, $d(x)=0$. Between the two intervals, the density interpolates continuously. As $|x|\rightarrow \infty$, $d(x)\rightarrow 1/2$.  The mirrored gray curve (plus the pale blue lines) corresponds to the density of ${\cal N}'_{AB}$ --- see \req{tomo} for definitions.}
	\label{dens}
\end{figure}

The modular flow for the chiral fermion was studied in several papers \cite{Casini:2009vk,Longo:2009mn,Longo:2017mbg,Hollands:2019hje}. In \cite{Hollands:2019hje}, the following useful correlator was computed for the case of a multi-interval region $(a_1,b_1)\cup\cdots \cup (a_k,b_k)$,
\be 
\langle \Omega|\psi(x)\Delta^{i t} \psi^\dagger(y)|\Omega \rangle=\frac{1}{2\pi i (x-y)} \frac{\Pi_b(x)\Pi_a(y)-\Pi_b(y)\Pi_a(x)}{e^{\pi t}\Pi_b(x)\Pi_a(y)-e^{-\pi t}\Pi_b(y)\Pi_a(x)}\,,
\ee
where
\be
\Pi_a(x)\equiv \prod_{i=1}^k (x-a_i)\,,\hspace{1cm} \Pi_b(x)\equiv \prod_{i=1}^k (x-b_i)\,.
\ee
Taking the limit $t\rightarrow -i/2$ from above, one finds\footnote{This can also be obtained from the explicit diagonalization of the modular operator \cite{Casini:2009vk}. According to (\ref{correla}) and (\ref{cij}) this is the analytic expression for $\sqrt{D(1-D)}(x,y)$ where $x$ is extended outside $AB$.}

\be 
\langle \Omega|\psi(x)\Delta^{1/2} \psi^\dagger(y)|\Omega \rangle=\frac{1}{2\pi  (x-y)} \frac{\Pi_b(x)\Pi_a(y)-\Pi_b(y)\Pi_a(x)}{\Pi_b(x)\Pi_a(y)+\Pi_b(y)\Pi_a(x)}+ i \,\pi\, g(x) \, \delta(y-\bar{y}(x))\, .\label{den}
\ee
The real part is given by the first term on the right hand side, and has a singular behavior in $y$ of the form
\be
\frac{g(x)}{y-\bar{y}(x)}\, ,
\ee
 near a point $\bar{y}(x)\in A$ given by the vanishing of the denominator in (\ref{den}). This defines $g(x)$. The imaginary part in the limit $t\rightarrow -i/2$ appears  associated to this same singular behavior of the real part, due to the Plemelj formula.

We get that the term proportional to a delta function in the  kernel (\ref{dos}) is
\be
4 \pi^2 \, g(x)g^*(x)\delta(\bar{y}(x)-\bar{y}(x')) \, ,
\ee
 and from this
 \be
d(x)=4\pi^2 \frac{|g(x)|^2}{|\bar{y}'(x)|}\, .
\ee 
Using \req{den}, it is possible to obtain an explicit expression for $d(x)$ in the case of two intervals. The resulting formula is a bit messy, but we can simplify it by considering two equal-size symmetric intervals: $a_1 \equiv -b$, $b_1\equiv-a$, $a_2\equiv a$, $b_2\equiv b$. In that case, we find
\begin{equation}
d(x)=\begin{dcases} \frac{(a^2(x^2-2b^2)+x(b^2 x + S(x)))^2}{2(ab +x^2) S(x) ((a-b)^2 x +S(x) )} \, , \quad &x\in [-a,a] \, ,\\     \frac{-(a^2(x^2-2b^2)+x(b^2 x - S(x)))^2}{2(ab +x^2) S(x) ((a-b)^2 x -S(x) )} 
\, , \quad &x \in (-\infty, -b] \cup [b,\infty)\, , \end{dcases}
 \end{equation}
 where
 \begin{align}
 S(x) &\equiv \sqrt{4a^3 b^3 +(a^4-4a^3 b-2a^2b^2-4 a b^3+b^4) x^2+4 a b x^4}\, .
 \end{align}
 We plot $d(x)$ in Fig. \ref{dens}. The red intervals correspond to $d(x)=1$ and $d(x)=0$, corresponding to $x\in A $ and  $x\in B$ respectively. As we can see, the type-I factor is spread through the whole line outside $B$. In particular, $d(x)$ asymptotes to $1/2$ as $|x| \rightarrow \infty$. The density for the factor ${\cal N}'_{AB}$ is the mirrored image of the one corresponding to ${\cal N}_{AB}$.

As a comparison, the type-III factor corresponding to interval $A$ has density equal to $1$ inside $A$ and $0$ outside. Then the continuous drop of the density outside $A$ is important for making the algebra type-I and have finite entropy. Note however that the density of ${\cal N}_{AB}$ is continuous and has continuous first derivative but does not have continuous second derivative. 

\section{Twist operators}\label{z2u1}
In this Section we first obtain explicit expressions, in terms of fermion correlators, for the expectation values on Gaussian states of twist operators  implementing global $\mathbb{Z}_2$ and $U(1)$ transformations. We evaluate those expectation values numerically for various angles as well as the``twist entropy'' defined in \req{ifo} for the $\mathbb{Z}_2$ case. We use this result to compute the type-I entropy defined in Section \ref{suba} for the bosonic subalgebra.

Let us consider a $U(1)$ symmetry group $g_\theta$ acting on the fermionic fields as
\begin{equation}
g_{\theta} \psi g_{\theta}^\dagger=e^{-i\theta}\psi\, .
\end{equation}
The twist operators $\tau_\theta$ act as the above group transformations for fields on $A$, while leaving the fields in $B$ invariant,
 \begin{equation}
\tau_{\theta} \psi(x) \tau_{\theta}^\dagger=e^{-i\theta}\psi(x)\,, \hspace{.5cm} x\in A\,,\hspace{1cm} \tau_{\theta} \psi(x) \tau_{\theta}^\dagger=\psi(x)\,, \hspace{.5cm} x\in B\,.\label{ddd}
\end{equation}
Then  $\tau_\theta$ is an operator localized in the complement of $B$.

Keeping only $\theta=0,\pi$, we restrict $U(1)$ to a $\mathbb{Z}_2$ group, 
 and we define $\tau\equiv \tau_{\pi}$, which satisfies $\tau^2=1$. This operator leaves invariant any product involving an even number of fermionic operators, while effectively multiplying by $-1$ an odd number of them, namely,
\begin{equation}
\tau \underbrace{\psi \cdots \psi}_{ n} \tau =(-1)^n \psi \cdots \psi \, .
\end{equation}

A would-be sharp twist with action (\ref{ddd}) but where $B$ is the complement $A'$ of $A$ does not correspond to any operator in the theory, because it would have too large fluctuations. This would correspond to a twist acting only on the type-III algebra which does not define a tensor product of the Hilbert space.
 This is not the case for type-I factors, for which $\tau_{\theta}$ is a well defined unitary.

 Consider the fermionic Gaussian state
\begin{equation}
\rho=\prod_l \frac{e^{-\varepsilon_l c_l^{\dagger}c_l}}{(1+e^{-\varepsilon_l})}=\otimes_l \left[\frac{\ket{0_l}\bra{0_l}+e^{-\varepsilon_l} \ket{1_l}\bra{1_l}}{(1+e^{-\varepsilon_l})}\right]\,,
\end{equation}
where the modular Hamiltonian has been diagonalized, as before. Considering the twist $\tau_\theta$ that acts on this Hilbert space, it follows that
\begin{equation}\label{tauuu}
\braket{\tau_{\theta}}=\tr (\rho \tau_{\theta})=\prod_l \frac{(1+e^{-(\varepsilon_l+i\theta)})}{(1+e^{-\varepsilon_l})}
\, ,
\end{equation}
where we defined the representation of the twist by $\tau_{\theta}= \otimes_{l} \tau_{\theta,l}$ and $\tau_{\theta,l} \ket{0_l}=\ket{0_l}$,   $\tau_{\theta,l} \ket{1_l}=e^{-i\theta}\ket{1_l}$. 
We can write \req{tauuu} in terms of the modular Hamiltonian as
\begin{equation}
\braket{\tau_{\theta}}=\det \left[ \frac{(1+e^{-(H+i\theta)})}{(1+e^{-H})} \right]\, .
\end{equation}
Finally, using the relation between the modular Hamiltonian and the Gaussian correlators $H=-\log(D^{-1}-1)$, this reduces to 
\begin{equation}
\braket{\tau_{\theta}}=\det \left[D+ (1-D) e^{-i \theta} \right]\, , \quad \braket{\tau}= \det [2D-1]\, .
\end{equation}

Just like for the reflected entropy, we can now apply these expressions in terms of the fermion correlator matrix to the type-I algebra $\mathcal{N}\equiv \mathcal{A}_A \vee J_{AB}   \mathcal{A}_A J_{AB}$, replacing $D$ by $C_{AA^*}$ above. This gives the expectation value of the standard twist defined by $A$ and $B$.

\begin{figure}[t]
	\centering 
	\includegraphics[scale=0.58]{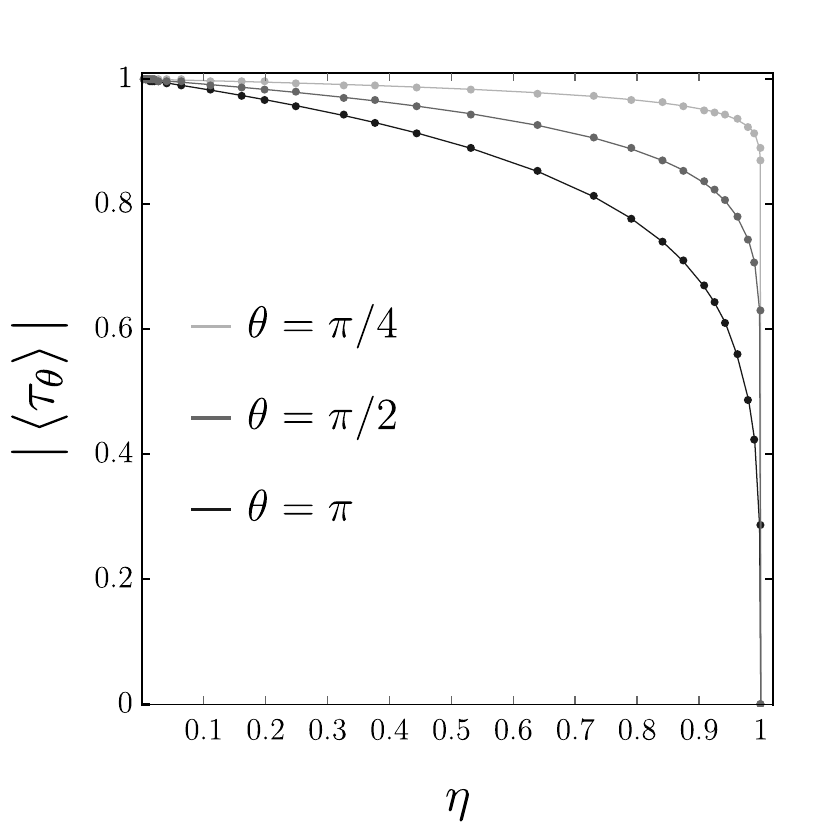}
	\includegraphics[scale=0.6]{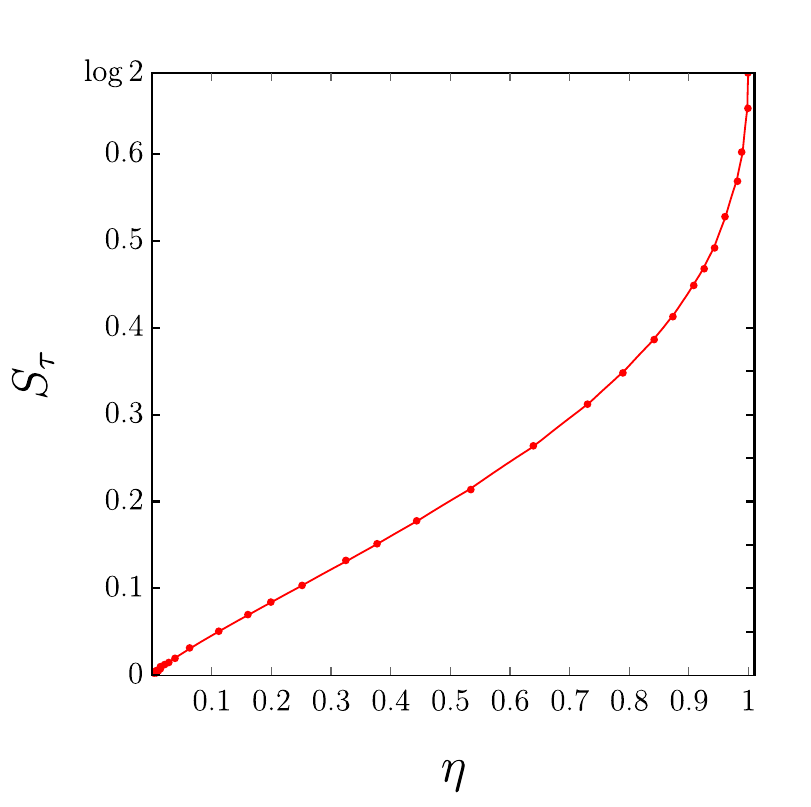}
	\caption{(Left) Expectation value of the standard twist operator $\tau_{\theta}$ for $\theta=\pi/4, \pi/2,\pi$ as a function of the cross-ratio $\eta$ for a free chiral fermion. (Right) Twist entropy $S_{\tau}$ associated to the $\mathbb{Z}_2$ symmetry, as defined in \req{z22}, for a free chiral fermion. The curve continuously grows from $S_{\tau}=0$ at $\eta=0$ to $S_{\tau}=\log 2$, its maximum value, at $\eta=1$.}
	\label{ratioss}
\end{figure}

We have computed these expectation values in the lattice and taken the continuum limit. 
In our lattice model with doubling we have two identical independent copies in the continuum limit. Then we have to take the square root of the lattice twist to get the expectation value of the twist in the chiral fermion right, $\langle \tau_\theta^{\textrm{lattice}}\rangle|_{\textrm{continuum}}=\langle \tau_\theta\rangle^2$. 
For two intervals the expectation value is a function of the cross ratio that we have plotted in Fig. \ref{ratioss} for $\theta=\pi,\pi/2,\pi/4$.
In the limit $\eta\rightarrow 1$ the twist is sharp, and charge fluctuations in vacuum make the expectation value go to zero. In the opposite limit, $\eta\rightarrow 0$, the twist smearing region between $A$ and $B$ becomes large, and the twist can transition smoothly between the group operation to the identity without appreciably disturbing the vacuum. In consequence, the expectation value $|\langle \tau_\theta\rangle|\sim 1$, as is the case of the expectation value of the group operation $\langle g_\theta \rangle=1$.    
 
 

\subsection{Type-I entropy for the bosonic subalgebra}

The bosonic subalgebra of the fermion model is defined by the collection of operators having even fermion number, or equivalently, operators invariant under the ${\mathbb{Z}_2}$ symmetry generated by $g_\pi$.\footnote{If we consider a real (Majorana) chiral fermion instead of a complex fermion, the bosonic model is the chiral Ising model (the Virasoro model of central charge $c=1/2$). The twists and type-I entropy can be studied in a similar way. In particular,  the expectation values of the twists are given by $\langle\tau_{\rm Ising}\rangle|=\sqrt{|\langle \tau \rangle|}$ because of the two independent real fermions in the complex one.}   
Using $\braket{\tau}$, we can compute the type-I entropy corresponding to the bosonic subalgebra,  $S^{{\rm I}}_{\rm bos.}(A,B)$. We have
\begin{equation}\label{siii}
 S^{{\rm I}}_{\rm bos.}(A,B)=R_{\rm ferm.}(A,B)-\frac{1}{2}S_{\tau}\,,
\end{equation}
where in this case the twist entropy is 
\begin{equation}\label{z22}
S_{\tau}=- \left[q_+ \log q_+ + q_- \log q_- \right]\, , \quad \text{where}\quad q_{\pm}\equiv \frac{1\pm \braket{\tau}}{2}\, .
\end{equation}
This follows from \req{mitad} and \req{qrr} as follows. First, $\mathbb{Z}_2$ has two irreducible representations, which we denote $r=+$ (trivial) and $r=-$, and two elements, $\{g_0,g_1\}$, whose characters read $\chi_+(g_0)=\chi_+(g_1)=1$, $\chi_-(g_0)=1$ and $\chi_-(g_1)=e^{i \pi}=-1$ respectively. As for the dimensions of the group and the two irreps, $|\mathbb{Z}^2|=2$, $d_+=1$ and $d_-=1$. Also, note that in the notation of \req{qrr}, we have $\braket{\tau_{g_0}}=1$, $\braket{\tau_{g_1}}=\braket{\tau}$. Inserting these results in \req{pro}, \req{qrr} and \req{mitad}, we are left with \req{z22}. In this case, the second term in the right hand side of \req{mitad} does not appear, since: $\log d_+=\log d_-=0$. The resulting curve for $S_{\tau}$ appears plotted in Fig. \ref{ratioss}.

\begin{figure}[t]
	\centering 
	\includegraphics[scale=0.67]{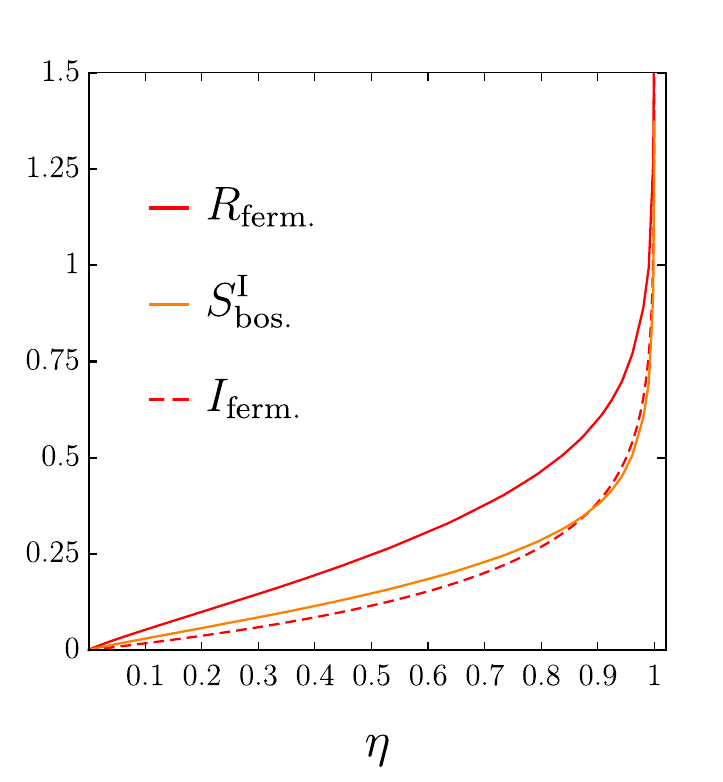}
	\includegraphics[scale=0.65]{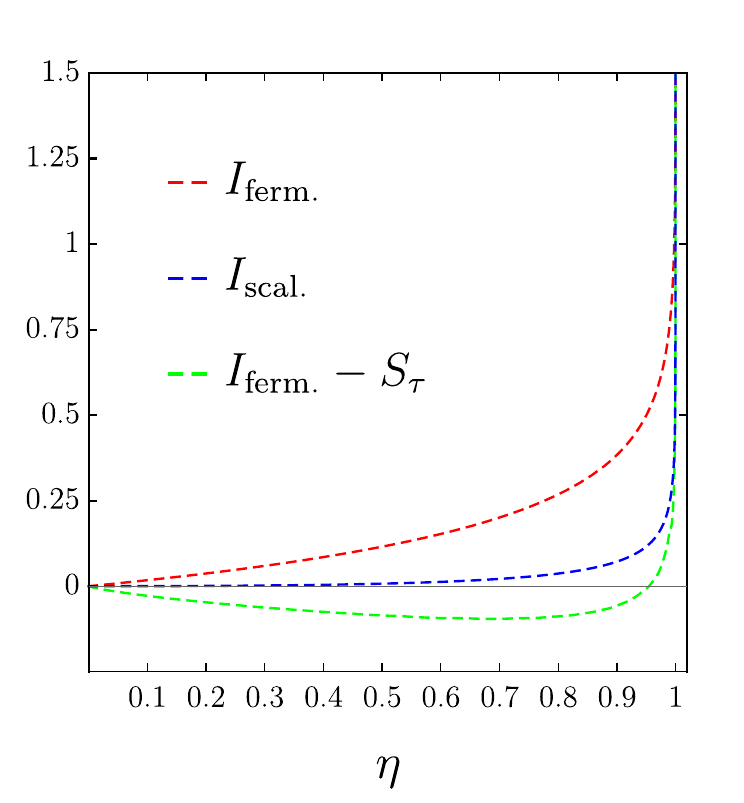}
	\caption{(Left) We plot the reflected entropy for the full fermion algebra, $R_{\rm ferm.}$, the type-I entropy for the bosonic subalgebra, $S^{\rm I}_{\rm bos.}$, and the mutual information for the fermion $I_{\rm ferm.}$ as a function of the cross-ratio $\eta$. (Right) We plot the free-fermion and free-scalar mutual informations, $I_{\rm ferm.}$ and $I_{\rm scal.}$ as well as $I_{\rm ferm.}-S_{\tau}$. }
	\label{refiss}
\end{figure}

Once we have $S_{\tau}(A,B)$, it is trivial to obtain $S^{{\rm I}}_{\rm bos.}(A,B)$ from \req{siii}. In Fig. \ref{refiss} we plot this ``type-I entropy'' alongside the reflected entropy of the full fermion algebra, $R_{\rm ferm.}(A,B)$, as well as the fermion mutual information, given by \req{mutuaf} above. We observe that 
\begin{equation}
R_{\rm ferm.}(\eta)> I_{\rm ferm.}(\eta)\, , \quad \text{and} \quad R_{\rm ferm.}(\eta)>S^{{\rm I}}_{\rm bos.}(\eta)\, ,
\end{equation}
for all values of $\eta$. These follow in general from \req{ini} and \req{siii} plus the positivity of $S_{\tau}$ respectively. On the other hand, we observe that $S^{{\rm I}}_{\rm bos.}(\eta)$ is quite close to $I_{\rm ferm.}(\eta)$ for all values of $\eta$. $S^{{\rm I}}_{\rm bos.}(\eta)$ is larger than $I_{\rm ferm.}(\eta)$ for smaller values of $\eta$, they coincide at some intermediate point $ \eta \sim  0.89$, and then $I_{\rm ferm.}(\eta)>S^{{\rm I}}_{\rm bos.}(\eta) $ as $\eta \rightarrow 1$. 

Using $S_{\tau}$ we can also obtain bounds for the mutual information of the bosonic subalgebra, $I_{\rm bos.}(\eta)$, which follow from  \req{ifo}. Namely, we have
\begin{equation}
 I_{\rm ferm.}(\eta)\geq I_{\rm bos.}(\eta)\geq I_{\rm ferm.}(\eta)-S_{\tau}(\eta)\, .
\end{equation}
$I_{\rm bos.}(\eta)$ is also bounded below by the mutual information of a free scalar field,
\begin{equation}\label{ibosc}
I_{\rm bos.}(\eta)\geq I_{\rm scal.}(\eta)\, .
\end{equation}
The last inequality follows from the monotonicity of mutual information under inclusions and the fact that the free scalar algebra is a subalgebra of the free-fermion bosonic one. Indeed, by bosonization, the free-scalar algebra is equivalent to the algebra generated by the fermion current, which includes only charge neutral operators. This includes smeared operators constructed from $\bar \psi( x)\psi(y)$ but not $\psi(x) \psi(y)$, for instance. This later however belongs to the bosonic subalgebra.

Note that writing a similar expression to \req{ibosc} for the reflected entropy is not possible at the moment, since we have no proof of the monotonicity of such quantity under inclusions.

\begin{figure}[t]
	\centering 
	\includegraphics[scale=0.76]{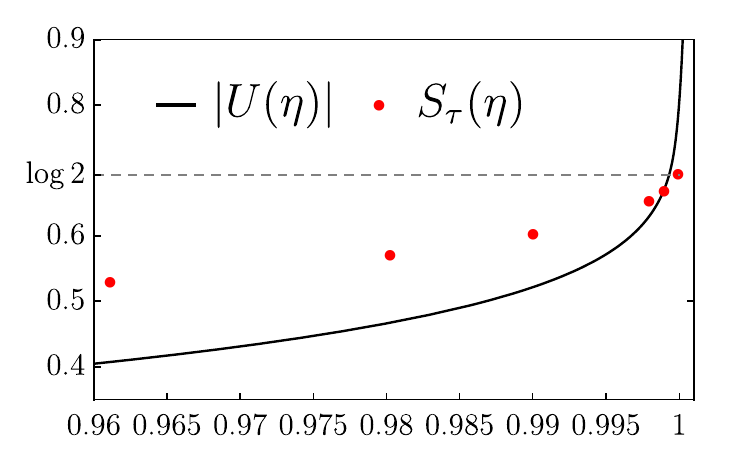}
		\caption{We plot $|U(\eta)|$ and $S_{\tau}(\eta)$ for values of the cross ratio close to $1$. $S_{\tau}(\eta)$ becomes smaller than $|U(\eta)|$ as we approach that limit, implying that $ I_{\rm ferm.}(\eta)-S_{\tau}(\eta)$ provides a better bound than $I_{\rm scal.}(\eta)$ to the bosonic subalgebra mutual information  near $\eta=1$.}
	\label{refiss2}
\end{figure}

The result for the mutual information of a free scalar reads \cite{Arias:2018tmw}
\begin{equation}
I_{\rm scal.}(\eta)=-\frac{1}{6}\log (1-\eta)+U(\eta)\, ,
\end{equation}
where
\begin{equation}
U(\eta)\equiv -\frac{i\pi}{2} \int_0^{\infty} ds \frac{s}{\sinh^2(\pi s)} \log \left[\frac{_2F_1[1+is,-is; 1; \eta]}{_2F_1[1-is,+is; 1; \eta]} \right] \, .
\end{equation}
We plot this together with $I_{\rm ferm.}(\eta)$ and $I_{\rm ferm.}(\eta)-S_{\tau}(\eta)$ in the second plot of Fig. \ref{refiss}. We observe that $I_{\rm scal.}(\eta)$ seems to be greater than  $I_{\rm ferm.}(\eta)-S_{\tau}(\eta)$ for all values of $\eta$, therefore providing a better bound for the mutual information of the bosonic subalgebra. In fact, $I_{\rm ferm.}(\eta)-S_{\tau}(\eta)$ turns out to be negative for most values of $\eta$. A closer look reveals that actually $I_{\rm ferm.}(\eta)-S_{\tau}(\eta)>I_{\rm scal.}(\eta)$ for values of $\eta$ sufficiently close to $1$. In that limit $I_{\rm ferm.}(\eta)-S_{\tau}(\eta)$ becomes positive, since $I_{\rm ferm.}(\eta)$ diverges whereas $S_{\tau}(\eta)$ approaches $\log 2$. The difference between the two quantities is given by
\begin{equation}
I_{\rm scal.}(\eta)-\left( I_{\rm ferm.}(\eta)-S_{\tau}(\eta)\right)=S_{\tau}(\eta)- |U(\eta)|\, ,
\end{equation}
where note that $U(\tau)$ is negative for all values of $\eta$. For most values of $\eta$, the above quantity is positive and therefore $I_{\rm scal.}(\eta)$ provides a better bound than $ I_{\rm ferm.}(\eta)-S_{\tau}(\eta)$. As it can be seen from Fig. \ref{refiss2}, this is no longer the case as $\eta\rightarrow 1$. $|U(\eta)|$ eventually becomes larger than $S_{\tau}(\eta)$ as we approach that limit, in fact becoming infinitely greater in the limit.

\section{Final comments}
The main results of the paper appear summarized in the introduction and at the beginning of each section. Let us now close with some final words.

In comparing the free-fermion reflected entropy with the mutual information, we have seen that both behave similarly as the intervals approach each other, but differ significantly otherwise. The reason for this difference is that mutual information measures correlations between operators strictly localized in $A,B$, while this is not the case of the reflected entropy. For the latter, information is more delocalized, as can be seen from the distribution of the type-I factor in space (see Fig. \ref{dens}). This is also manifest in the dependence of the reflected entropy on the cross ratio for large distances, which, as opposed to the mutual information, does not seem to have a power law expansion. In the case of the mutual information, such power law expansion follows from the OPE of localized twist operators in the replica trick \cite{Cardy:2013nua}. 

We have also computed explicitly some standard twist operators and their expectation values. These can be used to produce lower bounds on the mutual information of the orbifold theory and to compute the type-I entropy defined in Section \ref{suba}. The bound on the mutual information appears to be rather poor unless the two regions are near to each other. In the limit of regions touching each other the bound gets saturated. We can naturally wonder if there exist other twist operators, different from the standard ones, which produce sharper  bounds.  

Here we have focused on free fermions, but the reflected entropy for Gaussian bosonic systems should also be amenable to simple numerical study. Higher-dimensional studies of this quantity for free fields would of course be interesting as well. 
In particular, it would be interesting to analyze EE universal terms using reflected entropy as a regulator, and compare those with the mutual information regularization. In principle, we expect coincidence of results for the universal terms, as happens in the holographic case. Finally, let us mention that the holographic construction of reflected entropy \cite{Dutta:2019gen} may also give hints on how to compute standard twists in the holographic setup.

\section*{Acknowledgments}
We thank Tom Faulkner, Javier Mag\'an and Diego Pontello for useful comments. 
 This work was partially supported by the Simons foundation through the It From Qubit Simons collaboration. 
 H.C. was partially supported by CONICET, CNEA, and Universidad Nacional de Cuyo, Argentina.

\bibliographystyle{JHEP}
\bibliography{Gravities}

\newcommand{\noop}[1]{}

\providecommand{\href}[2]{#2}\begingroup\raggedright\begin{thebibliography}{10}

\bibitem{haag}
R.~Haag, \emph{Local Quantum Physics}.
\newblock Springer, 1992.

\bibitem{Witten:2018lha}
E.~Witten, \emph{{APS Medal for Exceptional Achievement in Research: Invited
  article on entanglement properties of quantum field theory}},
  \href{http://dx.doi.org/10.1103/RevModPhys.90.045003}{\emph{Rev. Mod. Phys.}
  {\bfseries 90} (2018) 045003},
  [\href{https://arxiv.org/abs/1803.04993}{{\ttfamily 1803.04993}}].

\bibitem{Casini:2006ws}
H.~Casini, \emph{{Mutual information challenges entropy bounds}},
  \href{http://dx.doi.org/10.1088/0264-9381/24/5/013}{\emph{Class. Quant.
  Grav.} {\bfseries 24} (2007) 1293--1302},
  [\href{https://arxiv.org/abs/gr-qc/0609126}{{\ttfamily gr-qc/0609126}}].

\bibitem{Casini:2015woa}
H.~Casini, M.~Huerta, R.~C. Myers and A.~Yale, \emph{{Mutual information and
  the F-theorem}}, \href{http://dx.doi.org/10.1007/JHEP10(2015)003}{\emph{JHEP}
  {\bfseries 10} (2015) 003},
  [\href{https://arxiv.org/abs/1506.06195}{{\ttfamily 1506.06195}}].

\bibitem{Buchholz:1973bk}
D.~Buchholz, \emph{{Product states for local algebras}},
  \href{http://dx.doi.org/10.1007/BF01646201}{\emph{Commun. Math. Phys.}
  {\bfseries 36} (1974) 287--304}.

\bibitem{Buchholz:1986dy}
D.~Buchholz and E.~H. Wichmann, \emph{{Causal Independence and the Energy Level
  Density of States in Local Quantum Field Theory}},
  \href{http://dx.doi.org/10.1007/BF01454978}{\emph{Commun. Math. Phys.}
  {\bfseries 106} (1986) 321}.

\bibitem{Doplicher:1982cv}
S.~Doplicher, \emph{{Local aspects of superselection rules}},
  \href{http://dx.doi.org/10.1007/BF02029134}{\emph{Commun. Math. Phys.}
  {\bfseries 85} (1982) 73--86}.

\bibitem{Doplicher:1984zz}
S.~Doplicher and R.~Longo, \emph{{Standard and split inclusions of von Neumann
  algebras}}, \href{http://dx.doi.org/10.1007/BF01388641}{\emph{Invent. Math.}
  {\bfseries 75} (1984) 493--536}.

\bibitem{Doplicher:1983if}
S.~Doplicher and R.~Longo, \emph{{Local aspects of superselection rules. II}},
  \href{http://dx.doi.org/10.1007/BF01213216}{\emph{Commun. Math. Phys.}
  {\bfseries 88} (1983) 399--409}.

\bibitem{Longo:2019pjj}
R.~{Longo} and F.~{Xu}, \emph{{Von Neumann Entropy in QFT}},
  \href{http://dx.doi.org/10.1007/s00220-020-03702-7}{\emph{Communications in
  Mathematical Physics} (Feb., 2020) },
  [\href{https://arxiv.org/abs/1911.09390}{{\ttfamily 1911.09390}}].

\bibitem{Narnhofer:2002ic}
H.~Narnhofer, \emph{{Entanglement, split and nuclearity in quantum field
  theory}}, \href{http://dx.doi.org/10.1016/S0034-4877(02)80048-9}{\emph{Rept.
  Math. Phys.} {\bfseries 50} (2002) 111--123}.

\bibitem{Otani:2017pmn}
Y.~Otani and Y.~Tanimoto, \emph{{Toward Entanglement Entropy with UV-Cutoff in
  Conformal Nets}},
  \href{http://dx.doi.org/10.1007/s00023-018-0671-9}{\emph{Annales Henri
  Poincare} {\bfseries 19} (2018) 1817--1842},
  [\href{https://arxiv.org/abs/1701.01186}{{\ttfamily 1701.01186}}].

\bibitem{Hollands:2017dov}
S.~Hollands and K.~Sanders, \emph{{Entanglement measures and their properties
  in quantum field theory}},
  \href{https://arxiv.org/abs/1702.04924}{{\ttfamily 1702.04924}}.

\bibitem{Dutta:2019gen}
S.~Dutta and T.~Faulkner, \emph{{A canonical purification for the entanglement
  wedge cross-section}},  \href{https://arxiv.org/abs/1905.00577}{{\ttfamily
  1905.00577}}.

\bibitem{Casini:2010nn}
H.~Casini, \emph{{Entropy inequalities from reflection positivity}},
  \href{http://dx.doi.org/10.1088/1742-5468/2010/08/P08019}{\emph{J. Stat.
  Mech.} {\bfseries 1008} (2010) P08019},
  [\href{https://arxiv.org/abs/1004.4599}{{\ttfamily 1004.4599}}].

\bibitem{Jeong:2019xdr}
H.-S. Jeong, K.-Y. Kim and M.~Nishida, \emph{{Reflected Entropy and
  Entanglement Wedge Cross Section with the First Order Correction}},
  \href{http://dx.doi.org/10.1007/JHEP12(2019)170}{\emph{JHEP} {\bfseries 12}
  (2019) 170}, [\href{https://arxiv.org/abs/1909.02806}{{\ttfamily
  1909.02806}}].

\bibitem{Kusuki:2019rbk}
Y.~Kusuki and K.~Tamaoka, \emph{{Dynamics of Entanglement Wedge Cross Section
  from Conformal Field Theories}},
  \href{https://arxiv.org/abs/1907.06646}{{\ttfamily 1907.06646}}.

\bibitem{Takayanagi:2017knl}
T.~Takayanagi and K.~Umemoto, \emph{{Entanglement of purification through
  holographic duality}},
  \href{http://dx.doi.org/10.1038/s41567-018-0075-2}{\emph{Nature Phys.}
  {\bfseries 14} (2018) 573--577},
  [\href{https://arxiv.org/abs/1708.09393}{{\ttfamily 1708.09393}}].

\bibitem{Nguyen:2017yqw}
P.~Nguyen, T.~Devakul, M.~G. Halbasch, M.~P. Zaletel and B.~Swingle,
  \emph{{Entanglement of purification: from spin chains to holography}},
  \href{http://dx.doi.org/10.1007/JHEP01(2018)098}{\emph{JHEP} {\bfseries 01}
  (2018) 098}, [\href{https://arxiv.org/abs/1709.07424}{{\ttfamily
  1709.07424}}].

\bibitem{Akers:2019gcv}
C.~Akers and P.~Rath, \emph{{Entanglement Wedge Cross Sections Require
  Tripartite Entanglement}},
  \href{https://arxiv.org/abs/1911.07852}{{\ttfamily 1911.07852}}.

\bibitem{Cui:2018dyq}
S.~X. {Cui}, P.~{Hayden}, T.~{He}, M.~{Headrick}, B.~{Stoica} and M.~{Walter},
  \emph{{Bit Threads and Holographic Monogamy}},
  \href{http://dx.doi.org/10.1007/s00220-019-03510-8}{\emph{Communications in
  Mathematical Physics} (July, 2019) 293},
  [\href{https://arxiv.org/abs/1808.05234}{{\ttfamily 1808.05234}}].

\bibitem{Kusuki:2019evw}
Y.~Kusuki and K.~Tamaoka, \emph{{Entanglement Wedge Cross Section from CFT:
  Dynamics of Local Operator Quench}},
  \href{http://dx.doi.org/10.1007/JHEP02(2020)017}{\emph{JHEP} {\bfseries 02}
  (2020) 017}, [\href{https://arxiv.org/abs/1909.06790}{{\ttfamily
  1909.06790}}].

\bibitem{Moosa:2020vcs}
M.~Moosa, \emph{{Time dependence of reflected entropy in conformal field
  theory}},  \href{https://arxiv.org/abs/2001.05969}{{\ttfamily 2001.05969}}.

\bibitem{Kudler-Flam:2020url}
J.~Kudler-Flam, Y.~Kusuki and S.~Ryu, \emph{{Correlation measures and the
  entanglement wedge cross-section after quantum quenches in two-dimensional
  conformal field theories}},
  \href{https://arxiv.org/abs/2001.05501}{{\ttfamily 2001.05501}}.

\bibitem{Bao:2019zqc}
N.~Bao and N.~Cheng, \emph{{Multipartite Reflected Entropy}},
  \href{http://dx.doi.org/10.1007/JHEP10(2019)102}{\emph{JHEP} {\bfseries 10}
  (2019) 102}, [\href{https://arxiv.org/abs/1909.03154}{{\ttfamily
  1909.03154}}].

\bibitem{Chu:2019etd}
J.~Chu, R.~Qi and Y.~Zhou, \emph{{Generalizations of Reflected Entropy and the
  Holographic Dual}},  \href{https://arxiv.org/abs/1909.10456}{{\ttfamily
  1909.10456}}.

\bibitem{Marolf:2019zoo}
D.~Marolf, \emph{{CFT sewing as the dual of AdS cut-and-paste}},
  \href{http://dx.doi.org/10.1007/JHEP02(2020)152}{\emph{JHEP} {\bfseries 02}
  (2020) 152}, [\href{https://arxiv.org/abs/1909.09330}{{\ttfamily
  1909.09330}}].

\bibitem{Buchholz:1985ii}
D.~Buchholz, S.~Doplicher and R.~Longo, \emph{{On Noether's Theorem in Quantum
  Field Theory}},
  \href{http://dx.doi.org/10.1016/0003-4916(86)90086-2}{\emph{Annals Phys.}
  {\bfseries 170} (1986) 1}.

\bibitem{Dijkgraaf:1989hb}
R.~Dijkgraaf, C.~Vafa, E.~P. Verlinde and H.~L. Verlinde, \emph{{The Operator
  Algebra of Orbifold Models}},
  \href{http://dx.doi.org/10.1007/BF01238812}{\emph{Commun. Math. Phys.}
  {\bfseries 123} (1989) 485}.

\bibitem{Casini:2019kex}
H.~Casini, M.~Huerta, J.~M. Magán and D.~Pontello, \emph{{Entanglement entropy
  and superselection sectors. Part I. Global symmetries}},
  \href{http://dx.doi.org/10.1007/JHEP02(2020)014}{\emph{JHEP} {\bfseries 02}
  (2020) 014}, [\href{https://arxiv.org/abs/1905.10487}{{\ttfamily
  1905.10487}}].

\bibitem{Longo:2017mbg}
R.~Longo and F.~Xu, \emph{{Relative Entropy in CFT}},
  \href{http://dx.doi.org/10.1016/j.aim.2018.08.015}{\emph{Adv. Math.}
  {\bfseries 337} (2018) 139--170},
  [\href{https://arxiv.org/abs/1712.07283}{{\ttfamily 1712.07283}}].

\bibitem{Longo:1989tt}
R.~Longo, \emph{{Index of subfactors and statistics of quantum fields. I}},
  \href{http://dx.doi.org/10.1007/BF02125124}{\emph{Commun. Math. Phys.}
  {\bfseries 126} (1989) 217--247}.

\bibitem{Doplicher:1969tk}
S.~Doplicher, R.~Haag and J.~E. Roberts, \emph{{Fields, observables and gauge
  transformations I}},
  \href{http://dx.doi.org/10.1007/BF01645267}{\emph{Commun. Math. Phys.}
  {\bfseries 13} (1969) 1--23}.

\bibitem{Casini:2009vk}
H.~Casini and M.~Huerta, \emph{{Reduced density matrix and internal dynamics
  for multicomponent regions}},
  \href{http://dx.doi.org/10.1088/0264-9381/26/18/185005}{\emph{Class. Quant.
  Grav.} {\bfseries 26} (2009) 185005},
  [\href{https://arxiv.org/abs/0903.5284}{{\ttfamily 0903.5284}}].

\bibitem{Casini:2005rm}
H.~Casini, C.~D. Fosco and M.~Huerta, \emph{{Entanglement and alpha entropies
  for a massive Dirac field in two dimensions}},
  \href{http://dx.doi.org/10.1088/1742-5468/2005/07/P07007}{\emph{J. Stat.
  Mech.} {\bfseries 0507} (2005) P07007},
  [\href{https://arxiv.org/abs/cond-mat/0505563}{{\ttfamily
  cond-mat/0505563}}].

\bibitem{Chung_2000}
M.-C. Chung and I.~Peschel, \emph{Density-matrix spectra for two-dimensional
  quantum systems},
  \href{http://dx.doi.org/10.1103/physrevb.62.4191}{\emph{Physical Review B}
  {\bfseries 62} (Aug, 2000) 4191–4193}.

\bibitem{Calabrese_2008}
P.~Calabrese and A.~Lefevre, \emph{Entanglement spectrum in one-dimensional
  systems}, \href{http://dx.doi.org/10.1103/physreva.78.032329}{\emph{Physical
  Review A} {\bfseries 78} (Sep, 2008) }.

\bibitem{2013PhRvB..87w5107L}
L.~{Lepori}, G.~{De Chiara} and A.~{Sanpera}, \emph{{Scaling of the
  entanglement spectrum near quantum phase transitions}},
  \href{http://dx.doi.org/10.1103/PhysRevB.87.235107}{\emph{Physical Review B}
  {\bfseries 87} (Jun, 2013) 235107},
  [\href{https://arxiv.org/abs/1302.5285}{{\ttfamily 1302.5285}}].

\bibitem{Laflorencie:2015eck}
N.~Laflorencie, \emph{{Quantum entanglement in condensed matter systems}},
  \href{http://dx.doi.org/10.1016/j.physrep.2016.06.008}{\emph{Phys. Rept.}
  {\bfseries 646} (2016) 1--59},
  [\href{https://arxiv.org/abs/1512.03388}{{\ttfamily 1512.03388}}].

\bibitem{Alba:2017bgn}
V.~Alba, P.~Calabrese and E.~Tonni, \emph{{Entanglement spectrum degeneracy and
  the Cardy formula in 1+1 dimensional conformal field theories}},
  \href{http://dx.doi.org/10.1088/1751-8121/aa9365}{\emph{J. Phys.} {\bfseries
  A51} (2018) 024001}, [\href{https://arxiv.org/abs/1707.07532}{{\ttfamily
  1707.07532}}].

\bibitem{Longo:2009mn}
R.~Longo, P.~Martinetti and K.-H. Rehren, \emph{{Geometric modular action for
  disjoint intervals and boundary conformal field theory}},
  \href{http://dx.doi.org/10.1142/S0129055X10003977}{\emph{Rev. Math. Phys.}
  {\bfseries 22} (2010) 331--354},
  [\href{https://arxiv.org/abs/0912.1106}{{\ttfamily 0912.1106}}].

\bibitem{Hollands:2019hje}
S.~Hollands, \emph{{On the modular operator of mutli-component regions in
  chiral CFT}},  \href{https://arxiv.org/abs/1904.08201}{{\ttfamily
  1904.08201}}.

\bibitem{Arias:2018tmw}
R.~E. Arias, H.~Casini, M.~Huerta and D.~Pontello, \emph{{Entropy and modular
  Hamiltonian for a free chiral scalar in two intervals}},
  \href{http://dx.doi.org/10.1103/PhysRevD.98.125008}{\emph{Phys. Rev.}
  {\bfseries D98} (2018) 125008},
  [\href{https://arxiv.org/abs/1809.00026}{{\ttfamily 1809.00026}}].

\bibitem{Cardy:2013nua}
J.~Cardy, \emph{{Some results on the mutual information of disjoint regions in
  higher dimensions}},
  \href{http://dx.doi.org/10.1088/1751-8113/46/28/285402}{\emph{J. Phys.}
  {\bfseries A46} (2013) 285402},
  [\href{https://arxiv.org/abs/1304.7985}{{\ttfamily 1304.7985}}].

\end{thebibliography}\endgroup

\end{document}